\documentclass[sn-mathphys,iicol]{sn-jnl}% Math and Physical Sciences 

\usepackage{amsmath}
\usepackage{bm}% bold math
\usepackage{graphicx}

\usepackage{float}
\usepackage{placeins}
\usepackage{caption}

\jyear{2022}%

\theoremstyle{thmstyleone}%

\theoremstyle{thmstyletwo}%

\theoremstyle{thmstylethree}%

\bibstyle{sn-mathphys}

\begin{document}

\title[Few-electron correlations after ultrafast photoemission from nanometric needle tips]{Few-electron correlations after ultrafast photoemission from nanometric needle tips}

\author*{\fnm{Stefan} \sur{Meier}}\email{stefan.m.meier@fau.de}
\equalcont{These authors contributed equally to this work.}
\author{\fnm{Jonas} \sur{Heimerl}}\email{jonas.heimerl@fau.de}
\equalcont{These authors contributed equally to this work.}

\author{\fnm{Peter} \sur{Hommelhoff}}\email{peter.hommelhoff@fau.de}

\affil*{\orgdiv{Department of Physics}, \orgname{Friedrich-Alexander-Universität Erlangen-Nürnberg (FAU)}, \orgaddress{\street{Staudtstraße 1}, \city{Erlangen}, \postcode{91058}, \country{Germany}}}

\abstract{
Free electrons are essential in such diverse applications as electron microscopes, accelerators, and photo-emission spectroscopy. Often, space charge effects of many electrons are a nuisance. Confined to extremely small space-time dimensions, even two electrons can interact strongly. In this case, the Coulomb repulsion can now be highly advantageous, because it leads to surprisingly powerful electron-electron correlations, as we demonstrate here. We show that femtosecond laser-emitted electrons from nanometric needle tips are highly anti-correlated in energy because of dynamic Coulomb repulsion, with a visibility of $56\,\%$. We extract a mean energy splitting of 3.3\,eV and a correlation decay time of 82\,fs. Importantly, the energy-filtered electrons display a sub-Poissonian number distribution with a second order correlation function as small as $g^{(2)} = 0.34$, implying that shot noise-reduced pulsed electron beams can be realized based on simple energy filtering. Even heralded electrons could become available for quantum-enhanced electron imaging protocols. Furthermore, we also reach the strong-field regime of laser-driven electron emission. We gain deep insights into how the electron correlations of the different electron classes (direct vs. rescattered) are influenced by the strong laser fields. Our work levels the field of quantum electron optics, with direct ramifications for shot noise-reduced and quantum electron imaging as well as direct measurements of correlated electrons from inside of strongly correlated matter. 
}

\maketitle

\begin{figure}[h!]
	\includegraphics[angle =-90,origin = c, width=1\linewidth]{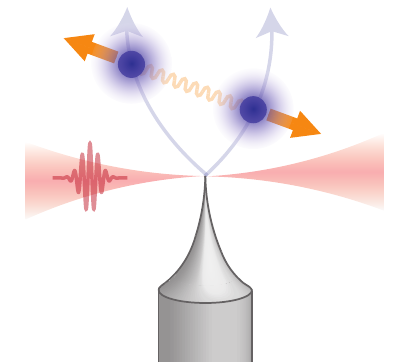}%
	\caption{Schematic of experiment. Electrons (blue) triggered by femtosecond laser pulses from a nanometric needle tip propagate towards a delay-line detector (not shown), which measures position and arrival time of each individual electron. Coulomb interactions (orange arrows) lead to a mutual repulsion of the electrons, altering trajectories (blue lines) and re-distributing kinetic energies of the two electrons. \label{fig:SetupFigure}}
\end{figure}

Electron correlations are central to intensely investigated cooperative effects inside of matter. The very nature and the relevant time scales of these effects bring together the fields of ultrafast physics and quantum (electron) optics \cite{Bloch2022}. The direct detection of two or more correlated electrons is highly sought after \cite{Kouzakov2003,Sobota2021}. Ideally, energy and momenta of the participating electrons could be measured directly. For this, photoemission spectroscopy with femtosecond time resolution lends itself ideal. Whereas the required photoemission setups with the proper detectors in principle exist \cite{Wang2013,Wallauer2021,Johnson2022}, it is yet imperative to first understand electron correlations arising dynamically after the electrons have left the sample (Fig. \ref{fig:SetupFigure}). In particular for imaging ultrafast effects from small volumes, these effects can quickly become dominant. We show here that even for electrons extremely confined in space-time to the nanometer-femtosecond range, two-particle effects are strong. Yet, the timescales are so fast that still a surprisingly large average current can be extracted before correlations due to Coulomb repulsion set in. Equally important, we show that these correlations in energy, in conjunction with an energy filter, can be used to attain electron beams with a sub-Poissonian counting statistics, highly relevant for imaging electron beam-sensitive specimen such as biological samples \cite{Glaeser2015}. Even experiments with heralded electrons are now conceivable, where the detection of one (energy-shifted) electron allows inferring that another electron must have interacted with the sample (Fig. \ref{fig:methods:heralding} in Methods). Sub-Poissonian electron sources can enhance the signal-to-noise ratio of any electron imaging device, and heralded electron sources may allow novel quantum imaging modes \cite{MagaaLoaiza2019}. To the best of our knowledge, our work shows for the first time strong energy correlations of electrons emitted from nanoscale solids - highly relevant for ultrafast electron beam applications including ultrafast electron microscopes \cite{Feist2017,Arbouet2018}, various kinds of time-resolved photoemission experiments \cite{Wang2013,Zong2018,Wallauer2021} and even nanophotonic particle accelerators \cite{England2014}. In close analogy to quantum optics, which deals with the particle properties of light and the quantum statistics of the photons, one may further argue that our work opens the field of quantum electron optics.

When the material under study is in the form of a nanometric needle tip, the electron source volume can be confined to extremely small length scales below  $\sim 10$\,nm. 
When the electrons are emitted with femtosecond laser pulses, an extremely high time resolution in the single-digit femtosecond scale can be achieved \cite{Ludwig2019,Hergert2021}. To elucidate the effects of this extreme space-time confinement, we combine ultrafast electron spectroscopy  with a multi-electron coincidence analysis on an event-by-event basis. We measure, for the first time, how Coulomb-induced correlations of two electrons after emission from the needle lead to a strong anticorrelation of the electrons’ energy, together with the associated time scale. 
Furthermore, by increasing the laser intensity and so entering the strong-field regime of photoemission, we observe how the electron anticorrelation diminishes. We can directly relate this to the interaction of different classes of electrons, namely electrons emitted without re-encounter with the parent tip (direct electrons) and electrons undergoing a laser field-induced scattering event with the tip after emission (re-colliding electrons). These insights represent the first step to investigate such intriguing effects as non-sequential double ionization from solids for the first time, so far only investigated from atoms and molecules \cite{lHuillier1983,Weber2000,Becker2012}.

In the experiment, we trigger electron emission from tungsten needle tips with radii of $10-15$\,nm by 12\,fs laser pulses at 800\,nm central wavelength, derived from an optical parametric amplifier with a repetition rate of 200\,kHz. We focus the laser pulses onto the tip by an off-axis parabolic mirror with a focal length of 15\,mm to a spot size of $\sim2\,\mu$m ($1/e^2$ intensity radius) (Fig. \ref{fig:SetupFigure}). The electrons are liberated from the metal tip by a nonlinear photoemission process in the transition region between multiphoton-photoemission and light-induced tunneling \cite{Hommelhoff2006_1,Ropers2007,Bormann2010,Krger2018}. They are further accelerated by a bias voltage of a few tens of Volts towards a delay line detector (DLD) with multi-hit capability. This detector allows us to measure the position and the time-of-flight (TOF) of one-, two- and three-electron events, for each electron individually. From the TOF and position, we calculate the energy of the electrons with an energy resolution of typically 0.3\,eV (for 40\,eV electrons). We magnify the central part of the electron beam by two quadrupoles with total magnification up to a factor of ten (see Methods).

We choose a mean electron count rate of 13.9\,kHz by setting the laser intensity to  $7.4\cdot10^{12}\,\mathrm{W}/\mathrm{cm}^2$. This way, we obtain a two-electron count rate of $\sim480\,$Hz, translating to $2.4\cdot10^{-3}$ double events per laser pulse, compared to only $1.4\cdot10^{-4}$ triple events. In total 3.5\,\% of all recorded events are two-electron events with both electrons recorded. We note that the given intensity already includes a field-enhancement factor of 3.7 obtained experimentally by an intensity sweep (see e.g. \cite{Thomas2013}).

\begin{figure*}[h!]
	\includegraphics[width=0.5\linewidth]{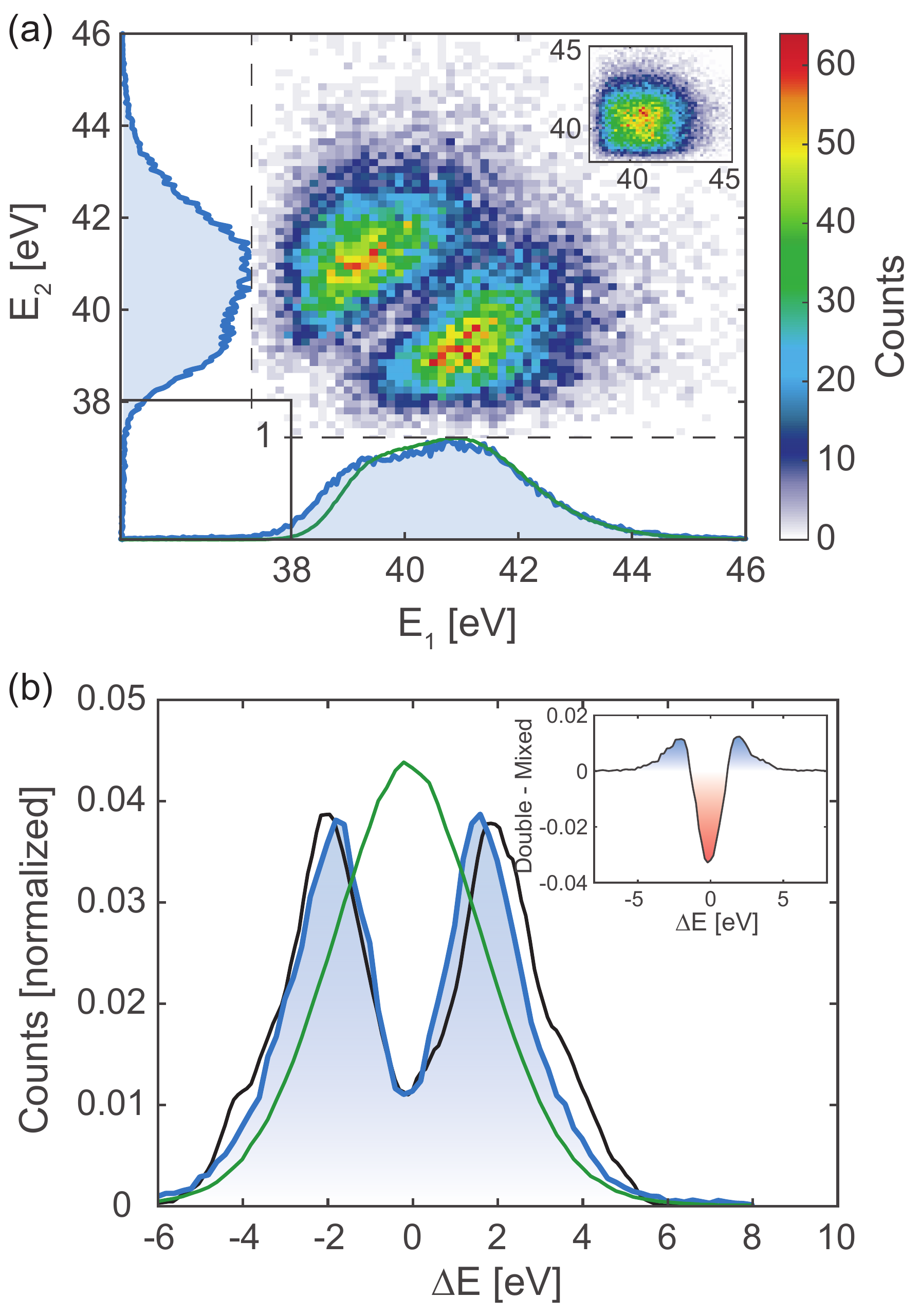}%
	\caption{Energy correlation in two-electron events. %Experimental data of Coulomb induced two-electron anticorrelation. 
	(a) Energy spectrum of the two-electron events: the energy of one electron is plotted over the energy of the other. A strong anticorrelation gap is observed along the diagonal, meaning that events with the same kinetic energy are strongly suppressed. The white line represents a 50\% contour line as a guide to the eye. The blue lines at the axes represent the energy spectra of only one and the other electron. In addition, the green line shows the single-electron spectrum for comparison. Clearly, the individual spectra of single and double electron events are very similar, and correlations only show up in the 2D representation between the double events. %All three energy histograms are normalized to their maximum. 
	The inset shows the corresponding map for two single electron events from two different laser pulses, with a clear maximum on the diagonal, as expected for uncorrelated events. (b) Same data plotted differently: histogram of the energy difference between the two electrons with a strong dip at zero energy difference, caused by Coulomb interactions between the two electrons after emission. For comparison, from arbitrary single events, we can generate uncorrelated double events [see the inset in (a)]. The green curve shows the energy difference of these events and shows no Coulomb-induced dip. A semi-classical simulation (black) quantitatively matches the experimental data. The inset shows the difference of the real double events and the mixed single events, both normalized to their respective number of events. Areas where there are more doubles than mixed singles are marked blue, vice versa red. \label{fig:ExperimentalGap}}
\end{figure*}

Measuring the energy for each of the two electrons yields the 2D energy correlation map shown in Fig.~\ref{fig:ExperimentalGap}(a): The horizontal axis shows the energy $E_1$ of one electron, the vertical the energy $E_2$ of the other, with arbitrarily chosen order. Clearly, two islands of high probability can be distinguished, one with a maximum at $E_1 = 41$\,eV and simultaneously $E_2 = 39$\,eV (and vice versa). At the respective axis, the energy spectra of only the one and only the other electron are shown (blue curves). On the horizontal axis, we further show a single-electron spectrum in green for comparison. So while the 1D-spectra give no hint of correlations, the 2D map shows clear evidence of an anticorrelation: two-electron events with the same energy, i.e. at the diagonal, are strongly suppressed and it is much more likely that electrons arrive with notably different energies. This anticorrelation is a signature of Coulomb repulsion between two emitted electrons within one laser pulse (in-depth discussion below). We note that we also see spatial electron correlations, which are harder to analyse due to specifics of our detector, and will need to remain for forthcoming work. 

A one-dimensional representation of the energy anticorrelation signal is given by the histogram of the energy difference \(\Delta E= E_2-E_1\) [Fig.~\ref{fig:ExperimentalGap}(b)]. We observe a repulsion visibility of $\mathcal{V}=56\,$\% of the dip, defined by \(\mathcal{V}=\frac{I_\text{max}-I_\text{min}}{I_\text{max}+I_\text{min}}\), where \(I_\text{max}\) is the maximum number of counts and \(I_\text{min}\) is the minimum number of counts around \(\Delta E=0\,\)eV. We extract the mean energy splitting as the gap size \(E_g\), given by the difference from peak to peak, equals \(E_g=3.3\)\,eV, notably much larger than the energy width of many electron beam devices like TEMs \cite{Jiang2018_2,Bach2019,Tsarev2021,Kuwahara2021}.

To gain quantitative insights and understand the origin of the anticorrelation in detail, we model our system with a semi-classical simulation: The emission process is treated quantum-mechanically, based on the emission of an electron in a laser field \cite{Yudin2001}. This provides the starting parameters for a Monte-Carlo simulation of the subsequent point-particle propagation (see Methods for details). With parameters matching the experiment, in particular a laser pulse duration of 12\,fs, a tip radius of \(r_\mathrm{tip}=15\,\)nm, an applied static field of 0.3\,V/nm and a laser intensity of $7.4\cdot10^{12}\,\mathrm{W}/\mathrm{cm}^2$, we obtain the black curve in Fig.~\ref{fig:ExperimentalGap}(b), almost perfectly matching the experimental data (see Methods). In the simulation, we assume a true random character of the emission. Hence, the excellent agreement between experiment and simulation shows that the Coulomb interaction between the two strongly space-time-confined electrons after the emission governs the spectra, as opposed to correlation effects in the emission process. This is corroborated by the scaling of the emission probabilities for 1-, 2- and 3-electron events (Methods Fig. \ref{fig:methods:Poisson_emisison}).

\begin{figure}[h!]
	\includegraphics[width=1\linewidth]{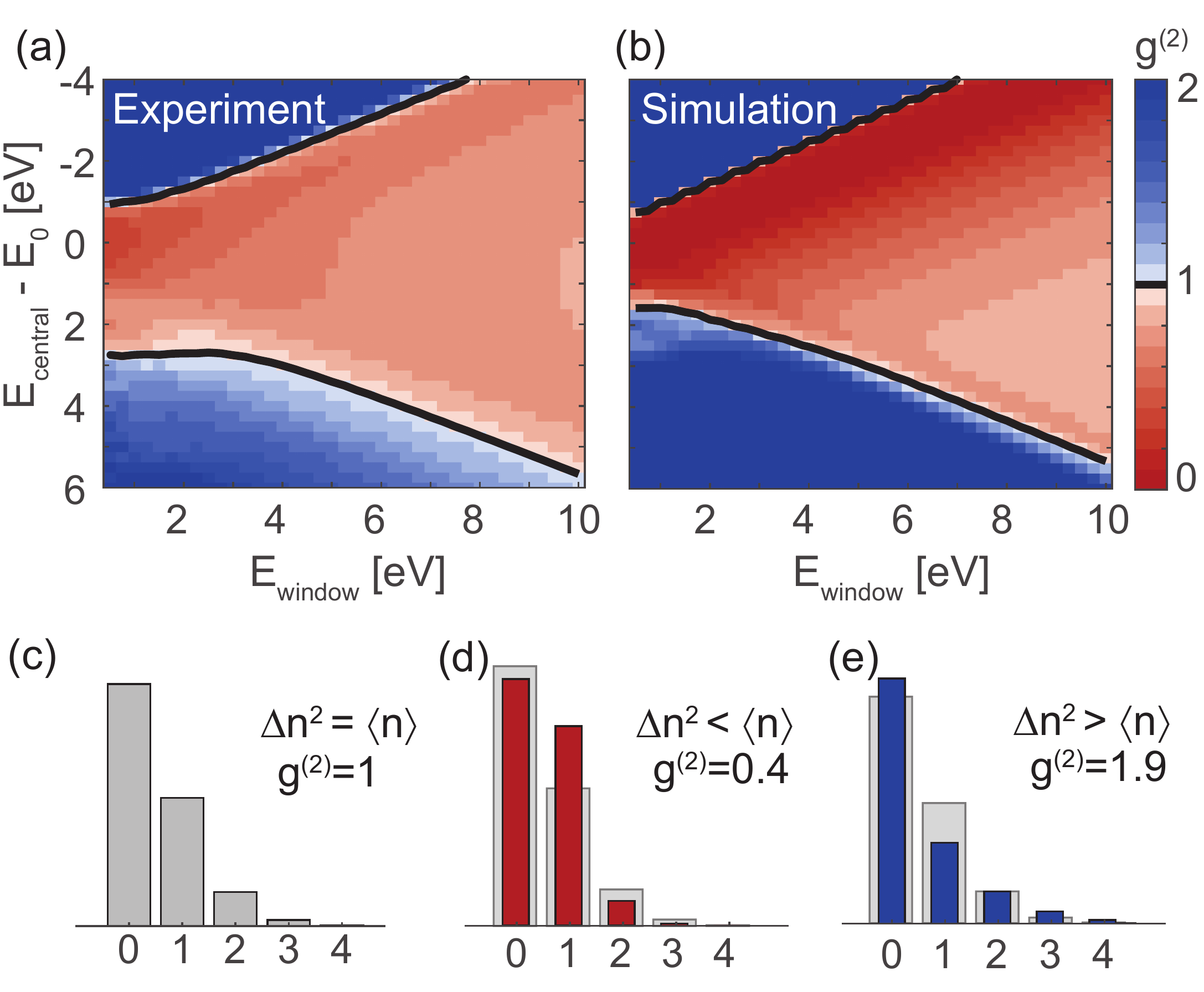}
    \caption{Second order correlation maps and electron number distributions. (a) Measurement of the second-order correlation function of a complete data set (measurement as in Fig. \ref{fig:ExperimentalGap}, only here with all events, not only the double-hits). $g^{(2)}$ was calculated for different subsequently applied energy filters. Filtering was done around a central energy $E_{\mathrm{central}}$ with a defined energy width $E_{\mathrm{window}}$. The mean energy $E_0$ of the single electron distribution was subtracted in the vertical axis. Depending on the filter settings, areas of lower variance (red, sub-Poissonian distribution) and higher variance (blue, super-Poissonian distribution) can be found. (b) Map of $g^{(2)}$ for a simulated data set, showing good agreement with (a). (c-e) show simulated electron number distributions. (c) Hit distribution for a Poisson distributed source with a mean value of 0.5 electrons per pulse. (d) Distribution for the same mean value after Coulomb interaction using a narrow-band energy filter. The variance of the hit distribution is smaller than its mean value, indicating a sub-Poissonian distribution. (e) Distribution using a broadband filter and higher central energy. Now a super-Poissonian distribution with a large variance is found.\label{fig:statistics}}
\end{figure}

The statistics of the detected electrons can be further quantified with the second order correlation function $g^{(2)}= \frac{\left<n(n-1)\right>}{\left< n \right>^2}$, where $n$ is the number of detected electrons within \(\sim 200\)\,ns, i.e., from within one laser pulse, and $\left< n \right>$ is the mean value of $n$. It is clear from Fig.~\ref{fig:ExperimentalGap}(a) that proper energy filtering can lead to sub- and super-Poissonian statistics: at the diagonal, the 2-electron coincidences show a dip, which is not the case for two uncorrelated single electrons, see inset. Thus, the resulting ratio around the diagonal between double- and single-events will become strongly reduced, leading to a reduced second order correlation function $g^{(2)}$ (see Method Fig. \ref{fig:methods:g2_map}). A similar effect has been observed with a two-pixel detector recently  \cite{Keramati2021}.

Fig. \ref{fig:statistics}(a) shows $g^{(2)}$ as function of different energy filter settings (see caption and Methods for details). We clearly find  sub-Poisson (red) and super-Poisson regions (blue). For comparison we show the same map generated from simulated data [Fig. \ref{fig:statistics}(b)]. Both maps show even quantitatively an almost identical behaviour. We find the minimum  $g^{(2)} = 0.34$ at the central energy of the electron distribution using a filter width of $0.5$\,eV.

For imaging applications, the Fano factor $F = \frac{\Delta n^2}{\left< n \right>}$ is of central importance, namely the ratio of variance and mean number of electrons per laser shot \cite{berchera2019quantum}. Immediately after emission our electron beam follows Poisson statistics where $F=1$. Dynamically, this evolves, after proper filtering, into a sub-Poissonian distribution, leading to a Fano factor of $F = 0.97\pm0.004$ in the experiment, close to the simulations results ($F=0.88$) with a mean of $0.5$ electrons per pulse (see Methods). Here, $F$ is only limited by the small average count rate necessary not to saturate our detector. Already with a mean number of 2 electrons per pulse, our simulations show that we can  achieve Fano factors down to 0.68 and up to 1.48, shown in the clear redistribution from Poisson to sub-Poisson and super-Poisson distributions [Fig. \ref{fig:statistics}(c-e)](see Methods). Sub-Poissonian electron beams are crucial for shot-noise reduced quantum imaging \cite{berchera2019quantum}, an extremely attractive feature for today's electron imaging application because of new single electron-counting detectors \cite{Jiang2018_2}.

\begin{figure*}[h!]
	\centering
	\includegraphics[width=1\linewidth]{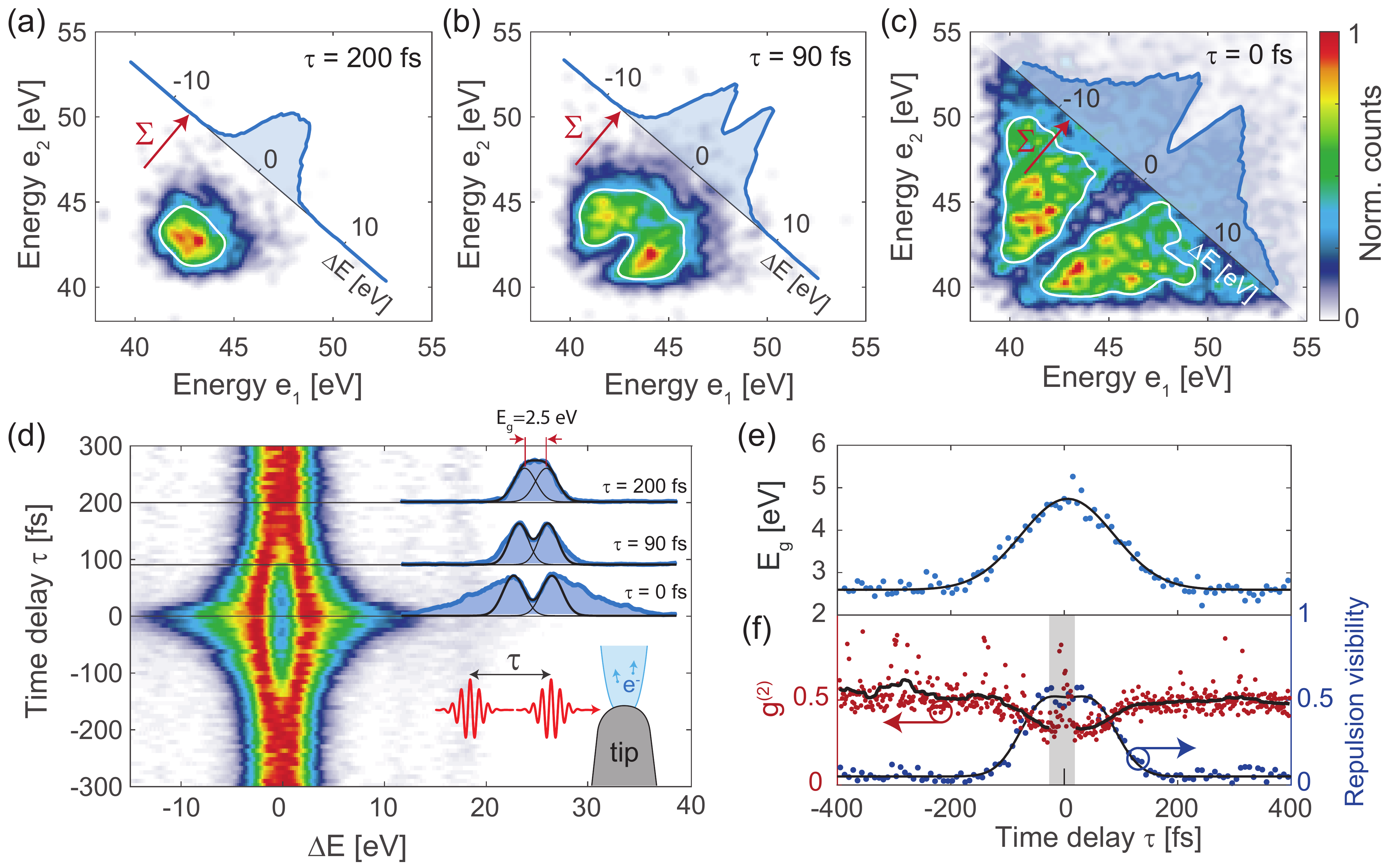}
	\caption{Ultrafast temporal behavior of the anticorrelations from a two-pulse measurement with adjustable time delay. (a-c) Energy correlation spectra for a pulse delay of \(200\,\)fs, \(90\,\)fs and \(0\,\)fs. From (a) to (c) one can clearly see an anticorrelation gap arising. Furthermore, the spectrum develops wings  because of the increased instantaneous laser power for smaller delays and resulting strongfield effects. The white lines represent 50\,\% contour-lines as a guide to the eye. The 1D-plots in (a-c) show respective energy difference spectra obtained by summation as indicated by the red arrows. (d) Energy difference spectra plotted for delays from \(-300\,\)fs to \(+300\,\)fs. A smooth transition from no gap around $\pm 200\,$fs to a pronounced gap around zero time delay can be clearly seen. The three energy difference spectra shown at (a-c) are plotted as insets at their respective time delay. A fit-function, given by the sum of two identical Gaussians (black lines) is shown for these three spectra as an example, see Methods for more details. (e) Energy gap size, (f) second order correlation function \(g^{(2)}\) (red) and repulsion visibility (blue) at \(\Delta E=0\,\)eV as function of pulse delay. A Gaussian function is fitted to the energy gap size, yielding  a standard deviation of \(\sigma_{\tau}=81.7\,\)fs. This correlation decay time corresponds to the average temporal separation of two electrons above which Coulomb interaction is suppressed. The black line above the data points of the \(g^{(2)}\) function is a moving-average function as guide to the eye. The filter for \(g^{(2)}\) is centered at \(E_0=43\)\,eV with 2\,eV width, at the minimum in the sub-Poissonian region (cf. Fig. \ref{fig:statistics}(a)). Note that for large delays \(g^{(2)}\) is still below one because of two electrons originating from one pulse only. The width of the repulsion visibility equals 69.1\,fs. The vertical gray bar indicates the region where strong-field effects affect both curves, see text for discussion.}
	\label{fig:figure2autocorrelation}
\end{figure*}

To obtain insights into the temporal behavior of the correlations, we focus a beam consisting of two copies of the same laser pulse with an adjustable time delay onto the tip to trigger the emission (intensity for each pulse: \(5.0\cdot10^{12}\,\mathrm{W}/\mathrm{cm}^2\)). In Fig. \ref{fig:figure2autocorrelation}(a-c), three electron energy correlation maps and their corresponding energy difference spectra are shown for three different time delays. For a time delay of \(\tau=200\,\)fs [Fig. \ref{fig:figure2autocorrelation}(a)], no anticorrelation gap is visible because the mean temporal separation is sufficiently large to prevent a significant Coulomb interaction. As the time difference of the laser pulses is decreased [\(\tau=90\,\)fs, Fig. \ref{fig:figure2autocorrelation}(b)], the anticorrelation gap emerges along the diagonal. For zero delay [Fig. \ref{fig:figure2autocorrelation}(c)], we again observe the clear gap, this time in conjunction with a broadened energy difference spectrum, which is due to the increased instantaneous intensity and resulting strong-field effects, see discussion below.

In Figure \ref{fig:figure2autocorrelation}(d) we show electron energy difference plots in the range of \(\tau=-300\,\mathrm{fs}\,\ldots+300\,\)fs in the form of a 2D map. The 2D map shows the anticorrelation gap as function of the delay between the two laser pulses: the gap smoothly opens up for time delays smaller than \(\sim170\,\)fs and vanishes for larger time delays. From this plot we can extract three characteristic features shown in Fig. \ref{fig:figure2autocorrelation}(e,f): (1) The width of the energy gap, extracted by fitting a double-Gaussian function to the energy differences [shown exemplary for three spectra in Fig. \ref{fig:figure2autocorrelation}(d)] as well as (2) the repulsion visibility and (3) $g^{(2)}$, all as function of pulse delay.

Due to the Poissonian distribution of the emission statistics, half of the two-electron events are composed of one electron triggered by the first and one by the second laser pulse (case 1), and the other half is triggered by one laser pulse only (case 2), independent of the total count rate. Therefore, only case-1 electrons show a delay-dependent energy gap, whereas case-2 electrons lead to a constant offset [Fig. \ref{fig:figure2autocorrelation}(e)]. This offset equals roughly half of the maximum energy gap at \(\tau=0\,\)\,fs, resulting from the fact that case-2 electrons are as many as case-1 electrons (see Methods).

By fitting a Gaussian function to the energy gap width, we infer the correlation decay time, i.e. the temporal range of the electron-electron interaction as \(\sigma_{\tau}=81.7\,\)fs [Fig. \ref{fig:figure2autocorrelation}(e)]. $g^{(2)}$ as function of the delay shows a similar temporal behavior with $\sigma_{\mathrm{\tau,g^{(2)}}} = 68.1\,$fs [Fig. \ref{fig:figure2autocorrelation}(f)],  coinciding with the value for the repulsion visibility $\sigma_{\mathrm{\tau,visibility}} = 69.1\,$fs. This timescale represents the emission time delay below which the electrons show a clear anticorrelation gap. Vice versa and intriguingly, this timescale  also yields a fundamental limit to the maximum emission rate of pulsed electrons of \(f_\mathrm{rep,max}=1/\sigma_{\tau}=12.2\,\)THz, the highest possible rate under which electrons with equidistant temporal spacing can be emitted from a nanometric tip without strong mutual interactions. The maximum current corresponding to \(f_\mathrm{rep,max}\) equals \(I_\mathrm{max}=e\cdot f_\mathrm{rep,max}=1.96\,\)µA, which is surprisingly large for most electron beam applications. It  represents a current limit for this type of nanometric electron emitter below which an undistorted electron beam can be achieved. Such a beam corresponds to a deterministic single electron source where in each emission event just one electron is emitted, with a constant temporal spacing. For a Poissonian-distributed emission process, such as standard laser-triggered electron emission or DC-field emission, this limit cannot be reached. However, pulsed laser-triggered deterministic single electron emitters are conceivable \cite{Zrenner2002,Hommelhoff2006_2,Duchet2021}, hence a deterministic high current single electron source seems within reach.

\begin{figure*}
	\centering
	\includegraphics[width=1\linewidth]{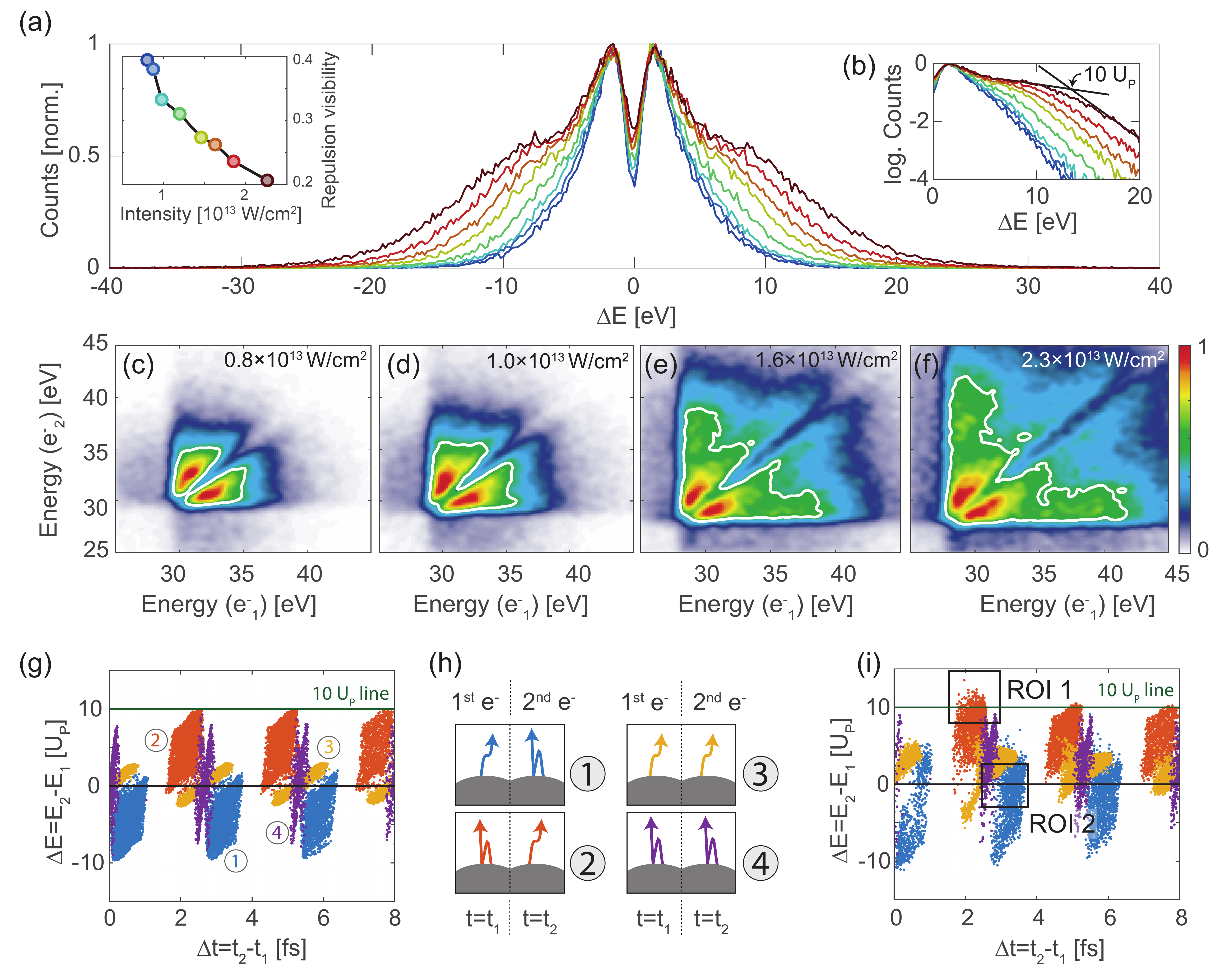}
	\caption{Laser intensity dependence of the observed anticorrelation gap and strong field effects. (a) Energy difference spectra as a function of incident laser intensity. The intensity steps are \(I_\mathrm{L}=(0.8,0.9,1.0,1.2,1.5,1.6,1.9,2.3)\cdot10^{13}\,\mathrm{W}/\mathrm{cm}^2\) (blue to brown lines). As the intensity increases, the gap becomes less pronounced. The repulsion visibility of the anticorrelation gap drops with increasing intensity, shown in the left inset. Further, a shoulder emerges, representing the famous rescattering plateau, visible well in the semi-logarithmic plot (b). The 10 \(U_\mathrm{P}\) cut-off for the highest intensity is at 13.3\,eV, in good agreement to the expected value of 13.5\,eV. (c-f) Electron correlation spectra showing the energy of one electron versus that of the other for four different laser intensities. % of \(I_\mathrm{L}=(0.8,1.0,1.6,2.3)\cdot10^{13}\,\mathrm{W}/\mathrm{cm}^2\). 
	The white 50\,\% contour lines show that the energy separation becomes suppressed for higher intensities. (g) Simulation results of the energy spectra as a function of start time difference \(\Delta t=t_2-t_1\) without Coulomb interaction for the four event classes shown in (h); the color code in (g) and (h) is matched. (h) Four event classes: \textbf{1}: the electron emitted first is a direct electron, leading to typically rather small final velocities, while the second is a rescattered electron with typically larger final velocities (resulting from strong-field dynamics in the large laser field \cite{Krger2018}); \textbf{2}: same as in \textbf{1} but roles swapped; \textbf{3}: both electrons are direct and \textbf{4}: both underwent rescattering. (i) like (h) but now with Coulomb repulsion between the electrons switched on. Clearly, all classes shift to larger \(\Delta E\). Class \textbf{2} now reaches final energies above \(10\,U_\mathrm{P}\), a region that cannot be reached for classical point-particles without interaction (see ROI 1). Both (g) and (i) were calculated for an intensity of \(1.8\cdot10^{13}\,\mathrm{W}/\mathrm{cm}^2\).}
	\label{fig:figure3powerdependence}
\end{figure*}

To investigate how strong-field effects \cite{Krger2011,Herink2012,Krger2018} affect the anticorrelation gap, we varied the incident intensity of (again single laser pulses) from \(8.0\cdot10^{12}\,\mathrm{W}/\mathrm{cm}^2\) to \(2.3\cdot10^{13}\,\mathrm{W}/\mathrm{cm}^2\), resulting in the energy difference spectra shown in Fig. \ref{fig:figure3powerdependence}(a). We observe that with increasing laser intensity, the gap depth at \(\Delta E=0\,\)eV becomes reduced, leading to a reduced repulsion visibility [inset in Fig. \ref{fig:figure3powerdependence}(a)]. Whereas the gap width stays almost constant, the individual width of the two peaks increases notably and a plateau arises (Fig. \ref{fig:figure3powerdependence}(a,b)). This plateau is the famous tell-tale feature of field driven dynamics: for certain emission times within the laser cycle the electron is driven back to the tip and scatters elastically off it \cite{Krger2011,Herink2012,Krger2018}. The maximum kinetic energy a rescattered electron can gain is ten times the ponderomotive energy (the famous 10 \(U_\mathrm{P}\) cut-off) \cite{Paulus1994}. This leads to an expected cut-off energy of 13.5\,eV for the highest intensity of \(2.3\cdot10^{13}\mathrm{W}/\mathrm{cm}^2\) in our measurement, which is nicely visible in Fig. \ref{fig:figure3powerdependence}(b).

Corresponding correlation maps for four different laser intensities are shown in Fig. \ref{fig:figure3powerdependence} (c-f). At the lowest intensity [Fig. \ref{fig:figure3powerdependence} (c)], the white 50\,\% contour line divides the spectrum in two parts. With increasing intensities both parts continuously grow together until the contour line as well as the highest count rate regions show almost no gap anymore [Fig. \ref{fig:figure3powerdependence} (f)]. Thus, the strong driving of the electrons in the laser field starts to mask the mutual Coulomb interaction, leading to a reduction of the anticorrelation gap.

To understand the correlated electron dynamics in the strong laser field, we carried out a 3D simulation including both strong field effects (three step model \cite{Corkum1993,Lewenstein1994}) and multi-electron Coulomb repulsion effects (see Methods). Fig. \ref{fig:figure3powerdependence}(g) shows the simulation results of the final energy difference of two electrons as function of their emission time difference with Coulomb interaction switched off. When we classify the strong-field emitted electrons into direct and rescattered electrons, we end up with four combinations [Fig. \ref{fig:figure3powerdependence}(h)]: (1) the first electron leaves the tip directly and the second undergoes rescattering, (2) the first electron undergoes rescattering whereas the second is emitted directly, and (3) \& (4), the two combinations where both electrons are direct and both undergo rescattering. These four classes are color-coded in Fig. \ref{fig:figure3powerdependence}(g), where the color matches the one in (h).

Around \(\Delta E=0\) and without Coulomb interaction, we mainly find events with two rescattered or two direct electrons. We repeated the simulation using the same starting conditions for each event but now turn on Coulomb interactions [Fig. \ref{fig:figure3powerdependence} (i)]. Whereas the four events classes are slightly fuzzier, most prominently they are shifted with respect to the case without Coulomb interactions. We now define two regions of interest (ROI) to analyze the main experimental features, namely the broadened spectrum and the reduced gap depth. From Fig. \ref{fig:figure3powerdependence}(i) we see that if the first electron is a rescattered one (high final energy) followed by a direct one (smaller energy), these events result in high energy differences (ROI 1): The trailing slow direct electron pushes the fast rescattered electron to even higher energies, also above the \(10\,U_\mathrm{P}\) line, the (classical) maximum energy for a single electron. In contrast, if a slow direct electron is followed by a fast rescattered one, the first electron slows down the second, faster one, hence both electrons end up with similar energies and thus around \(\Delta E=0\,\)eV, see ROI 2. This qualitatively explains our experimental findings in the inset of Fig. \ref{fig:figure3powerdependence}(a): We observe a decrease of the contrast when increasing the intensity since the influence of the laser field on the final energy of the electrons increases with increasing intensity. 

With this microscopic explanation we can quantitatively understand how the spectra evolve as a combination of (single electron) strong-field emission physics and (two or multi-electron) Coulomb interactions. Besides the decrease in repulsion visibility (ROI2), we see that electrons show up at higher energies than expected from single-particle models, evidenced by ROI1. We infer that to observe the interaction between two electrons in the clearest way, it is beneficial to work with the lowest laser intensity possible. For our experiments, the intensity required for a high-contrast gap is typically below \(\sim1\cdot10^{13}\,\,\mathrm{W}/\mathrm{cm}^2\). This marks an important difference between our tip-based experiments and electron-electron interaction in nonsequential double-ionization at atoms, where the typical intensity to observe two electron events is \(>1\cdot10^{14}\,\,\mathrm{W}/\mathrm{cm}^2\) \cite{Weber2000,Camus2012}. We suggest that the higher intensity required for atom-based correlation experiments is one of the main reasons why the observed anticorrelation gap is so much more pronounced in our case of needle-tip emission; in atoms, the anticorrelation gap like shown in Fig. \ref{fig:ExperimentalGap}(b) is virtually invisible \cite{Rudenko2007,Staudte2007,Becker2012}.

\begin{figure}[h!]
    \centering
    \includegraphics[width = 1\linewidth]{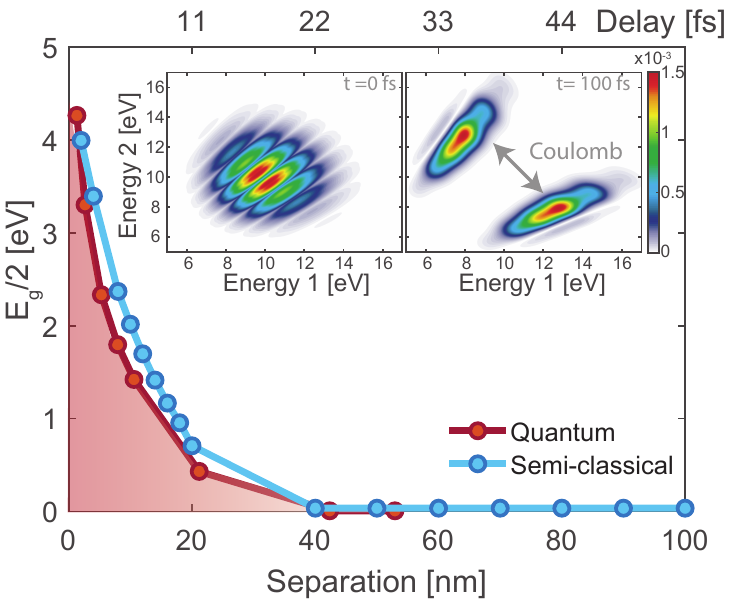}
    \caption{Energy gap size from quantum and semi-classical 1D simulations. The electrons are modelled as wave packages with a mean kinetic energy of 10\,eV and an energy spread of 2\,eV. Both simulations show the same behavior for the Coulomb-induced energy separation as function of the initial wave packet separation. Inset: The square modulus of the two-electron wavefunction $\left\vert\psi(E_1,E_2)\right\vert^2$ in energy space for an initial separation of 5\,nm, after 0\,fs (left) and 100\,fs (right) of propagation. Because of overlap in space and energy at 0\,fs we observe clear interference, the signature of quantum statistical effects. This interference vanishes completely  after 100\,fs because of the separation of  electrons in both spatial and energy domain due to the Coulomb interaction. See text for details.}
    \label{fig:Figure4}
\end{figure}

Due to the semiclassical nature of our simulations, quantum statistical effects were so far not included, such as the Pauli exclusion principle. Yet, the probability of electrons arriving at zero delay can be further suppressed by the Pauli exclusion principle \cite{Kiesel2002,Kodama2011,Kuwahara2021}. Thus, the Pauli principle could add to the energy gap in our case as well. To estimate its strength, we numerically solved the one-dimensional time-dependent Schrödinger equation for two electrons with Coulomb interaction (see Methods). The two insets in Fig. \ref{fig:Figure4} show the square modulus of the two-electron wavefunction in energy space right after emission ($t=0\,$fs) and after 100\,fs propagation. Because of spatial and energetic overlap we observe interference effects at $t=0\,$fs. After 100\,fs, the two-electron wavefunction shows a strong suppression along the diagonal. This energy separation demonstrates that quantum statistic effects cannot play any role after this time scale, as both spatial and energetic overlap would be required. When we vary the initial separation of the two wavepackets in the quantum simulation, we can extract the Coulomb-induced energy gap size after 100\,fs of propagation (red curve in Fig. \ref{fig:Figure4}). For comparison, we carried out a semi-classical one-dimensional simulation using the same initial starting conditions (blue curve). Since both curves show an almost perfect overlap, we conclude that the electrons in our case mainly behave like classical point-particles following the center of mass of their wavepackets, justifying our semi-classical modeling approach. Only when the kinetic energy is much higher and the initial temporal spread is larger, it is possible to clearly observe Pauli suppression, as vacuum dispersion and Coulomb interactions are reduced so that the wavefunction overlap during detection is much larger  \cite{Kiesel2002,Kodama2011,Kuwahara2021}.

In conclusion, we observed a strong anticorrelation behavior in two-electron emission from metal needle tips, clearly identified as Coulomb interaction arising dynamically after emission. This repulsion between both electrons leads to a prominent energy splitting with a visibility of up to \(\mathcal{V}=56\,\)\% and a gap width of \(\sim3.3\,\)eV, much larger than the electron beam spectral width in standard TEMs \cite{Jiang2018_2,Bach2019,Tsarev2021,Kuwahara2021}, for example. We observe $g^{(2)} = 0.34$, hence a sub-Poissonian pulsed electron beam is straightforwardly attainable by energy filtering. We show that the anticorrelation is most prominent when the emission times of the two electrons are separated by less than \(\sim80\)\,fs, which we identify as the correlation decay time. This timescale gives an upper current limit for a possibly unperturbed electron beam of \(\sim 2\,\)µA. For the highest intensities in the experiment, strong-field effects lead to a less pronounced gap, as well as electrons reaching energies above \(10\,U_\mathrm{P}\), which we fully understand. The experimental data are backed by semiclassical as well as 1D quantum-mechanical simulations.

Our work introduces correlation measurements into ultrafast electron emission from solids. We foresee our work to herald quantum-enhanced electron imaging modes, opening the field of quantum electron optics. Furthermore, when examining strong electron correlations in direct two electron emission experiments from solids, such as superconductors, the always present Coulomb interaction of the freed electrons could easily mask quantum-cooperative effects. From our measured timescale we can estimate the required lifetime of a correlated electron system to be probed in a fashion unperturbed from dynamically arising Coulomb interactions, which is as short as \(\sim80\,\)fs, depending on the correlation energy. 

In the final phase of manuscript writing we became aware of similar work by Rudolf Haindl, Armin Feist, Till Domröse, Marcel Möller, Sergey V. Yalunin, and Claus Ropers (see current submission to Nature Physics).

\begin{appendices}

\newpage
\FloatBarrier
%TC:ignore
\section*{Methods}\label{secA1}
\setcounter{figure}{0} 
\renewcommand{\figurename}{Extended Data Fig.}

\subsection*{Experimental details}
\paragraph{Tip handling and vacuum system}
Our experiments are carried out in an ultra-high vacuum chamber with a pressure \(<10^{-9}\,\)mbar. The vacuum chamber is mounted on an damped optical table as well as our laser system to avoid vibrations. The [310] tungsten tips are inserted into the chamber via a load-lock system. The tips are in-situ characterized by field ion microscopy (FIM) and cleaned in the same step by field evaporation.

\paragraph{Laser system and optics}
We use 12-femtosecond pulses from a commercial optical parametric chirped-pulse amplifier (OPCPA) from Laser Quantum. It runs at a repetition rate of 200 kHz at a central wavelength of 800 nm. We use chirped mirrors and fused silica wedges for dispersion compensation. The pulses are fed into the vacuum chamber, where they are focused down with an off-axis parabolic mirror with a focal length of 15 mm.\\
For the two-pulse measurement, we use a dispersion balanced Mach-Zehnder-interferometer to generate two copies of the laser pulse with an intensity ratio of 1 to 1.

\begin{figure*}
	\centering
	\includegraphics[width=1\linewidth]{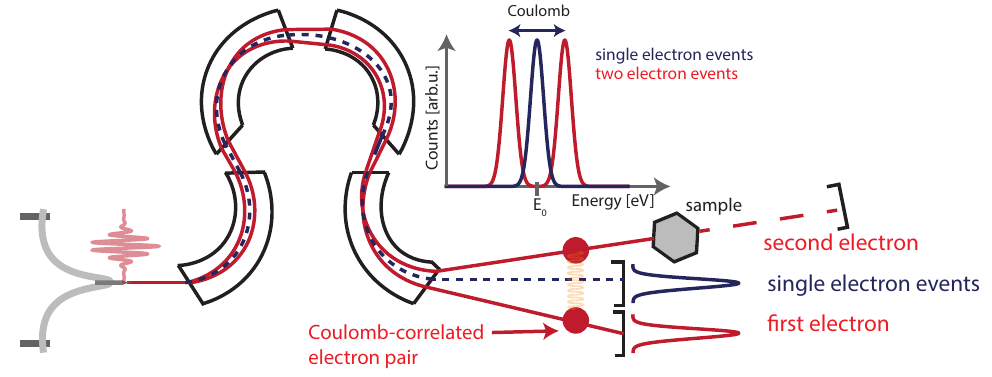}
	\caption{Measurement scheme for heralded of electrons. Electron events triggered from a metal needle tip are energetically separated by a Omega filter. When the energy width of single electron events is smaller than the mean Coulomb energy splitting, the electrons can be separated after the omega filter. Single and two-electron events are then separated in absolute energy. The measurement of one electron in the energy region prohibited for single electron events, only possible for two-electron events, leads directly to the knowledge of the presence of a second electron. By post-selection a deterministic electron source is achieved, which can be used for quantum imaging.
	\label{fig:methods:heralding}}
\end{figure*}

\subsection*{Electron detection system}
The electrons are detected by a delay-line detector, that consists of two multi-channel plates (MCP) with a diameter of 80 mm followed by three delay-line anodes (Hex-detector from RoentDek). Due to the redundant information of the third layer, this detector has multi-hit capability. For each particle, the position in x- and y- coordinate as well as the time of flight (TOF) is evaluated. From these coordinates we can calculate the particles momenta \(p_x=m_e\cdot x/(\mathrm{TOF}-t_0), p_y=m_e\cdot y/(\mathrm{TOF}-t_0)\) and \(p_z=m_e\cdot L/(\mathrm{TOF}-t_0)\), with the distance \(L\) between tip and detector and the electron mass \(m_e\). The time off-set \(t_0\) is governed by a fit to different electron spectra depending on the acceleration voltage. The kinetic energy is given by
\begin{align*}
E_\mathrm{kin}=\frac{\left(p_x^2+p_y^2+p_z^2\right)}{2m_e}
\end{align*}
for each particle. The energy resolution mainly depends on the mean kinetic energy, the temporal resolution \(\Delta t\) of the hardware and the distance to the detector. It is given by \cite{Damm2011_phd}:
\begin{align*}
	\Delta E=\sqrt{\frac{8E_\mathrm{kin}^3}{m_0}}\frac{\Delta t}{L}.
\end{align*}
For a temporal resolution of \(\Delta t=250\,\)ps, \(L=26\,\)cm and a mean kinetic energy of \(E_\mathrm{kin}=40\,\)eV, we obtain \(\Delta E\approx0.3\,\)eV.\\
As every detector based on delay lines, our detector has a certain space-time dead radius \cite{Jagutzki2002_2}. When two simultaneous hits are close, the signals at the delay lines start to merge. The precision of the reconstruction of position and time of each electron depends crucially on the number of obtained delay-line signals. To reduce the problem of the dead radius, while still having access to close electron events, we use two electrostatic quadrupoles on a motorized translation stage. These electrostatic lenses allow us to magnify the lateral electron beam radius up to a factor of 10, zooming into the central region of the electron beam. This way, we make sure not to count any detector artifacts as physical correlations. We determine the zoom factor by placing a TEM-grid close to the tip apex, generating a shadow image with and without quadrupoles.

Last, we note that the front MCP is a funnel-type MCP with a high quantum efficiency of 86\,\% \cite{Fehre2018}. Therefore we neglect quantum efficiency in our evaluation, however, we will address possible effects in future work on sub-Poissonian statistics.

\begin{figure*}
	\centering
	\includegraphics[width=0.8\linewidth]{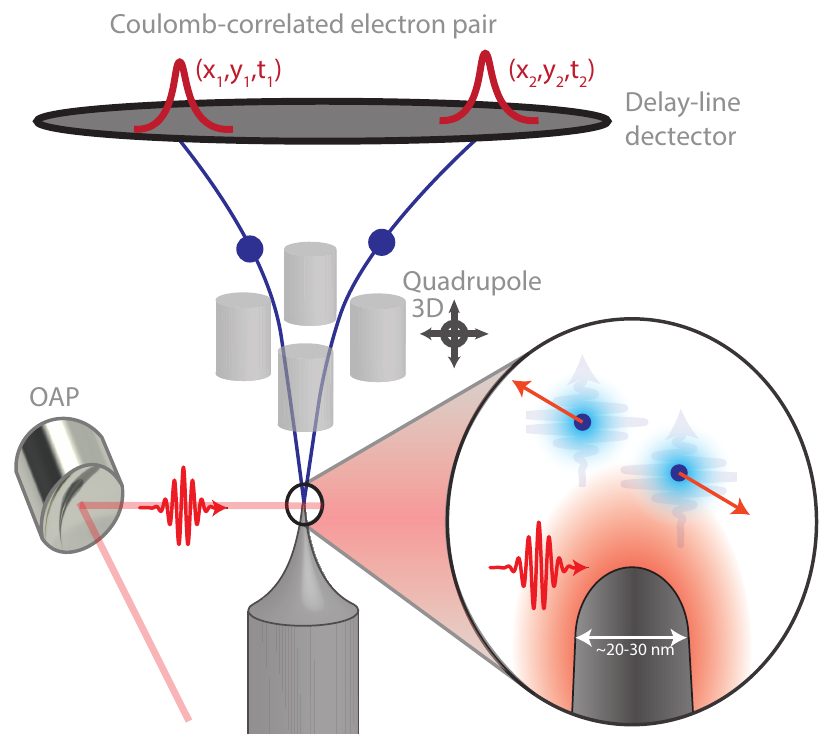}
	\caption{Sketch of experimental setup. Femtosecond laser pulses focused by an off-axis parabolic mirror (OAP) trigger electrons (blue) from a metal needle tip. The highly coherent electron beam is magnified by two quadrupoles by a factor of up to 10 (only one quadrupole shown here). The two quadrupoles can be moved by a 3-axis manipulation stage. The multi-hit capability of the delay-line detector allows us to measure the position x and y, and the time of flight for each electron individually. \label{fig:methods:setup}}
\end{figure*}

\subsection*{Two-pulse measurement}

We obtain the energy difference by fitting the sum of two Gaussian functions to the data in each line:
\begin{align*}
    f(x)=b_1\cdot\left[\exp\left(\frac{-(x+b_2)^2}{b_3^2}\right)+\exp\left(\frac{-(x-b_2)^2}{b_3^2}\right)\right]
\end{align*}
with fitting parameters \(b_1,b_2,b_3\). The parameter \(b_2\) equals half of the the energy width \(E_g=2\cdot b_2\) in our definition.

We note that in this measurement we observe a larger energy gap at \(\tau=0\)\,fs as compared to the gap shown in Fig. \ref{fig:ExperimentalGap}(a). We suspect that this is because the quadrupole settings were different in this case, resulting in a different magnification. Similarly, the exact shape and state of the needle tip can add small changes to the shape of the repulsion gap.

\subsection*{3D point-particle simulation}
For the simulation, we use a self-written point-particle trajectory simulation in \textsc{Matlab}. This semi-classical simulation uses a quantum mechanical emission probability function, calculated following \cite{Yudin2001}. After emission, we numerically integrate the equations of motion including the static electric field resulting from the bias voltage at the tip, the oscillating laser field and Coulomb interactions between the electrons. The equation of motion of one electron with index \(i\) is given by:
\begin{align*}
\ddot{\vec{r_i}}=\frac{e}{m_e}\vec{E}_i(\vec{r}_1,\dots,\vec{r}_n)
\end{align*}
with the elementary charge \(e\). The total electrical field is given as 
\begin{align*}
\vec{E}_{i}&=-\sum_{j\not= i}\frac{e}{4\pi\varepsilon_0}\frac{\vec{r}_i-\vec{r}_j}{\left\vert\vec{r_i}-\vec{r_j}\right\vert^3}+\vec{E}_\text{static}(\vec{r}_i)+\vec{E}_\text{laser}(\vec{r}_i),
\end{align*}
here \(\varepsilon_0\) is the vacuum permittivity and \(\vec{r}_{i,j}\) are the coordinates of two different particles \(i\) and \(j\). For \(E_\mathrm{static}\), we use the static field of a spherical capacitor, as we model the tip as a sphere (that is emitting in the positive half space) with the spherical counter electrode set to infinity:
\begin{align*}
	&\vec{E}_\text{static}(\vec{r}_i)=\\
	&=U_\mathrm{tip}\cdot\frac{1}{r_\mathrm{tip}+2(\vert\vec{r}_i\vert-r_\mathrm{tip})+\frac{(\vert\vec{r}_i\vert-r_\mathrm{tip})^2}{r_\mathrm{tip}}}\cdot\frac{\vec{r_i}}{\vert\vec{r}_i\vert},
\end{align*}
where \(U_\mathrm{tip}\) is the applied static voltage and \(r_\mathrm{tip}\) the tip radius. The laser field is approximated by a cosine with a Gaussian envelope:
\begin{align*}
E_\text{laser}(\vec{r}_i)=\gamma_\mathrm{nf}(\vec{r}_i)E_0\exp\left(-2\ln(2)\frac{t^2}{\tau^2}\right)\cos(\omega t)
\end{align*}
Here, \(\gamma_\mathrm{nf}(\vec{r}_i)=1+(\gamma_0-1)e^{-\vert\vec{r_i}\vert/r_\mathrm{tip}}\) is the near field decay with \(r_\mathrm{tip}\) as the characteristic decay length \cite{Seiffert2018}. In our simulation, the strength of the effective optical near field is the projection on the axis parallel to the tip´s shank. Satisfying Maxwell´s equation the field vector point orthogonal away from the surface, i.e. radially at the tip apex.

For the simulation in Fig. \ref{fig:ExperimentalGap}(b) we smoothed the distribution of high energies to compensate for the known artifacts of the simple-man`s model.

\subsection*{The three step model}
The three-step model was initially invented as a theoretical model to describe the strong-field electron emission and dynamics at atoms, but can also be applied to metallic needle tips \cite{Krger2012_2}. The three-step model consists of the following steps: first, one, two or even more electrons are emitted from the tip's surface during a negative half-cycle of the optical field. In the second step, they are driven in the laser field and are accelerated back to the surface by the subsequent positive half-cycle. In step three, the electrons rescatter elastically at the surface and are called rescattered electrons. If an electron does not reach the surface in the second step, it does not rescatter and is thus called a direct electron, as usual. At each time step, the instantaneous acceleration of the electrons is calculated as a function of the static field at the tip, the laser field, and the Coulomb interaction with all other emitted electrons.

\subsection*{Emission statistics}
In Fig. \ref{fig:methods:statistics}(a) we show the method how we generate the 2D second order correlation maps in Fig. \ref{fig:methods:statistics}(b), depicted also in the main text. We choose a central filter energy $E_{\mathrm{central}}$ around which only electrons within $\pm E_{\mathrm{window}}/2$ are taken into account. While single and double electron events can be precisely reconstructed, events with more than two electrons are very difficult to disentangle. To avoid the influence of falsely reconstructed electron events, we focus for the sub-/super Poisson measurements on count rates where the unfiltered distribution has a small ratio of events with more than two electrons (\(1.6\cdot 10^{-3}\) \%). The shown measurement has an average of \(3.3\cdot10^{-2}\) electrons per pulse, resulting in $g^{(2)} =  1.04$ and $F = 1.00$ without filtering, being an almost perfect Poisson distribution. We demonstrate by simulation that smaller Fano factors can be achieved by increasing the average count rate, as shown in Fig. \ref{fig:methods:statistics}(c). The horizontal axis is the mean number of totally emitted electrons, and the vertical axis the determined Fano factor after filtering at the detector plane. The individual histograms (orange) represent the electron number distribution for each simulated data point. The smaller the Fano factor gets, the more the difference to a Poisson distribution becomes apparent. In comparison, we show Poissonian distributions in gray in the background for the corresponding mean particle number after filtering. Already for an average of five emitted electrons per pulse at the tip, we observe a distribution with a Fano factor of $F = \sim 0.5$ around one electron per pulse on average after filtering. Going to even higher count rates, represents a road map for achieving small Fano factors, important for imaging.

For the experimentally determined Fano factors, we estimate the error bar by splitting our data set into four independent sub-sets. For each sub-set we calculate the Fano factor and then the standard deviation of the four obtained values, resulting in $\Delta F  = \pm 0.004$.

 \begin{figure*}
	\centering
	\includegraphics[width=0.8\linewidth]{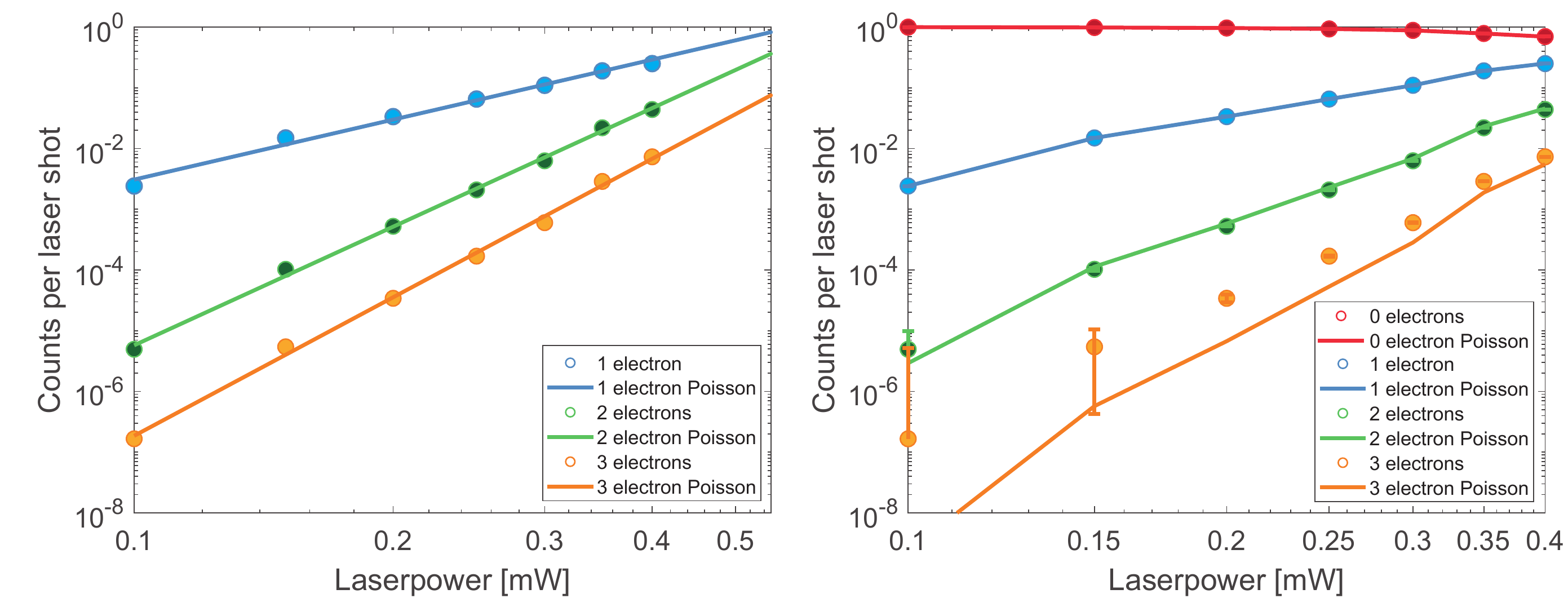}
	\caption{(a) Power scaling of n-electron events. Because of the multi-photon photoemission process, the electron emission follows a power law, visible by the linear scaling in the double-logarithmic representation. The slopes are $m=3.3\pm0.4$ (one electron), $m=6.5\pm0.4$ (two electrons) and $7.6\pm0.4$ (three electrons). The slope for the total emission (sum of all events) is $m = 3.4\pm 0.4$. (b) Same data as in (a). Solid lines show the theoretical Poisson distribution for n-electron events calculated from the measured mean of electrons per laser shot. The error bars are determined from the dark count rate of the detector. \label{fig:methods:Poisson_emisison}}
\end{figure*}

\begin{figure*}
	\centering
	\includegraphics[width=1\linewidth]{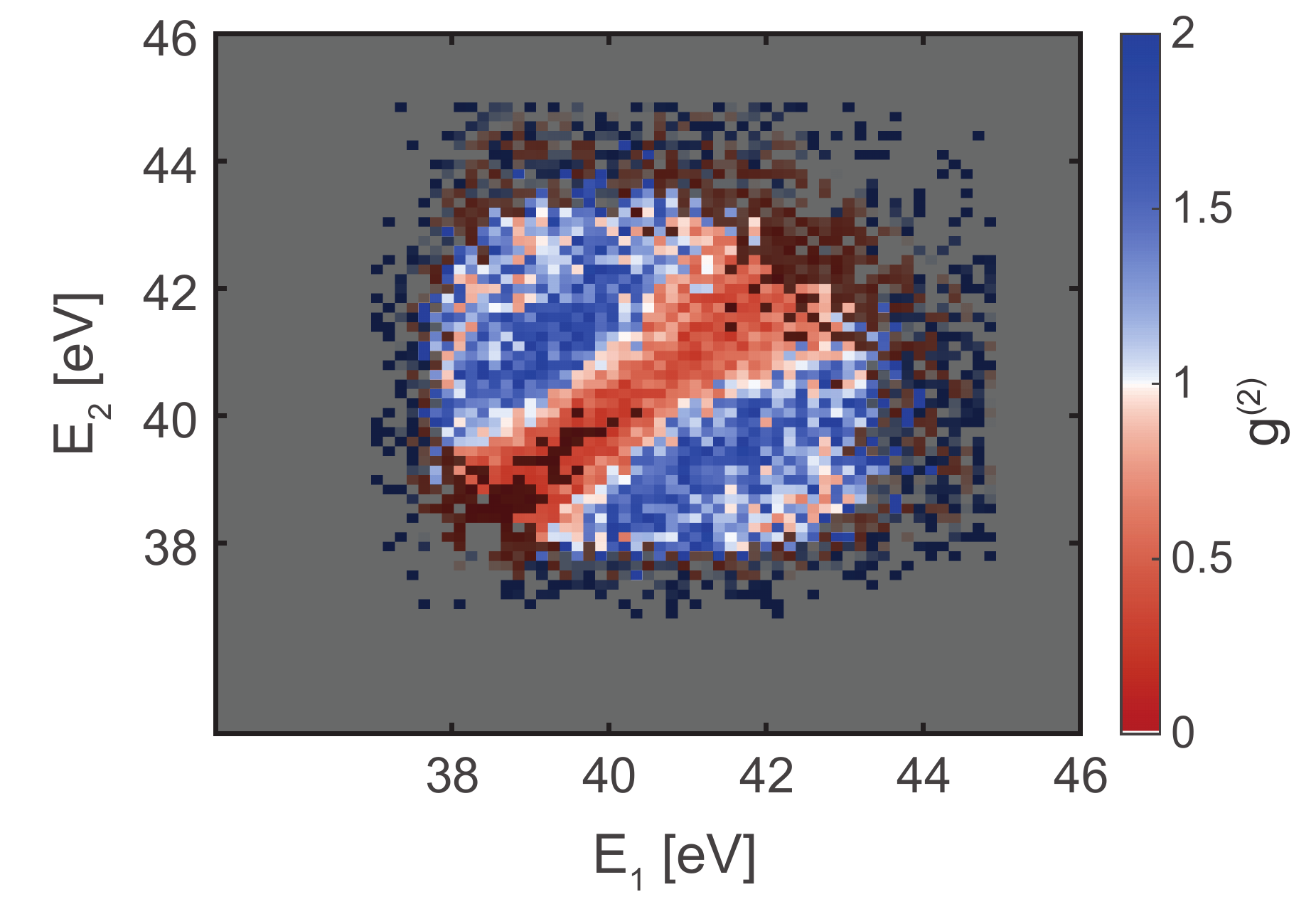}
	\caption{Second order correlation function $g^{(2)}(E_1,E_2) = \frac{\left< I(E_1)I(E_2)\right>}{\left<I(E_1)\right>\left<I(E_2)\right>}$ calculated for the energy map shown in Fig. \ref{fig:ExperimentalGap}(a). A filter excluding data points with less than 5 counts per bin removes data points dominated by noise (gray mask). \label{fig:methods:g2_map}}
\end{figure*}

\begin{figure*}
	\centering
	\includegraphics[width=1\linewidth]{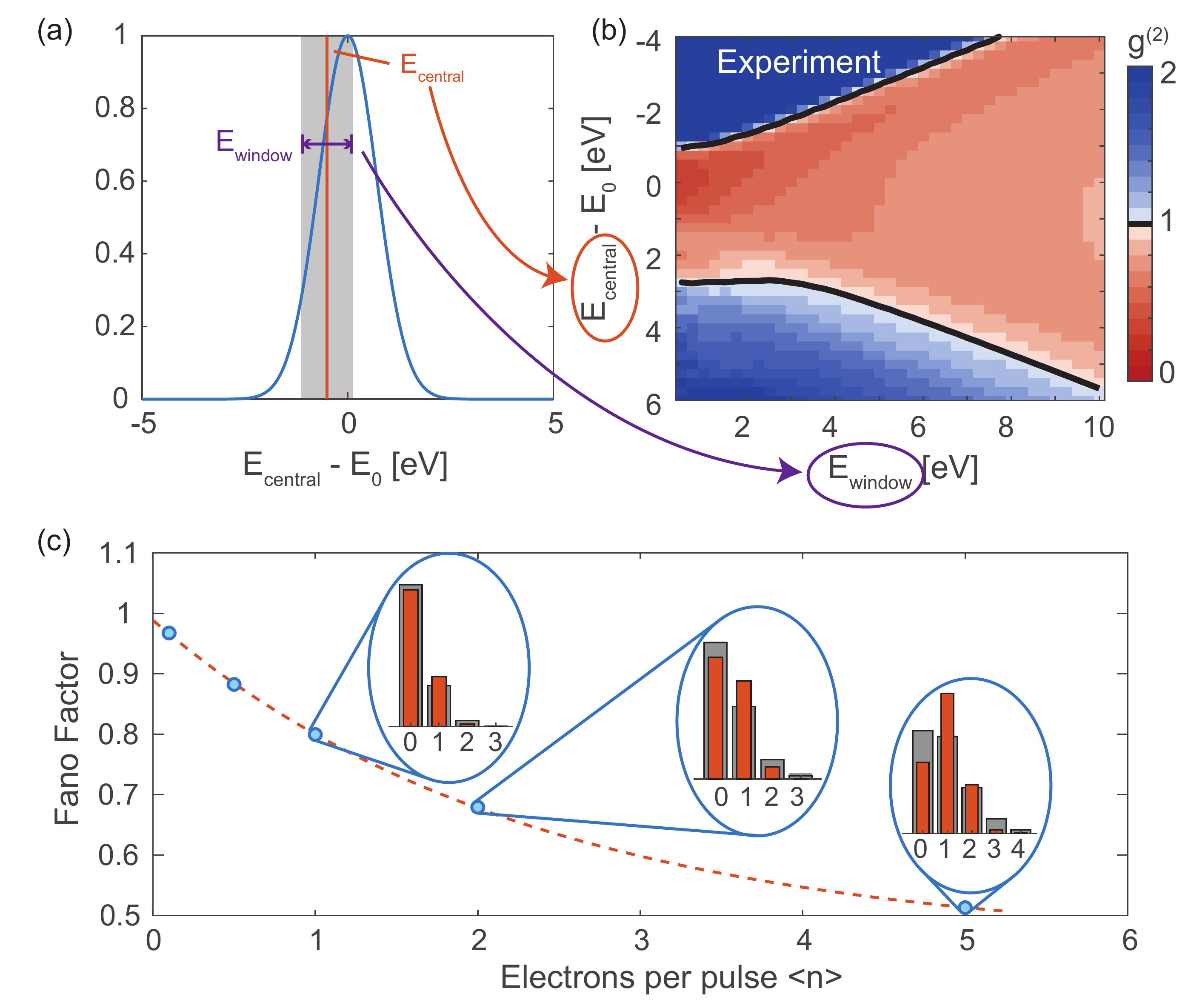}
	\caption{(a) Explanation of the energy filter used in Fig. \ref{fig:statistics}: a central energy is chosen (orange line) and then an interval of \(\pm E_\mathrm{window}/2\) is used as filter area. (b) Experimental data shown in Fig.\ref{fig:statistics} (a), with highlighted variables. (c) Fano factor for simulations from an average of 0.1 electron per pulse to 5 electrons per pulse. The insets show the resulting energy-filtered hit-distributions after propagation including Coulomb interactions. \label{fig:methods:statistics}}
\end{figure*}

\subsection*{Two-particle TDSE simulation}

Contrary, to the standard example of quantum mechanics textbooks, we are interested in the interaction of not only one but two particles. Following the approach of \cite{Maestri2000}, we numerically solve the two-particle time-dependent Schroedinger equation
\begin{align*}
    i\frac{\partial}{\partial t}\psi(x_1,x_2,t) = H\psi(x_1,x_2,t),
\end{align*}
where the Hamilton in atomic units is given by 
\begin{align*}
H = -\frac{1}{2}\frac{\partial^2}{\partial x_1^2}-\frac{1}{2}\frac{\partial^2}{\partial x_1^2}+V(x_1,x_2).
\end{align*}
In our case we use a soft Coulomb-potential
\begin{align*}
    V(x_1,x_2) = 1/(\lvert x_1-x_2 \rvert+a_0)
\end{align*} in order to avoid singularities at $x_1 = x_2$. We set $a_0 = 1$, which is sufficiently small so that the Coulomb-potential looks similar to the classical one over most of the region of our grid. In this calculation we do not consider static and laser fields.
The two wave packets are chosen to have similar properties to the experiment, being 10\,eV start energy and 2 eV energy spread. For the numerical propagation a trade-off between computational time, numerical stability and propagation time is required. Because of the dispersion of the wave packets over time, we need a sufficiently large grid (~700-1000\,\r{A}) while the spacing ($\textless 1$\,\r{A}) must be so small that a large enough k-space is covered, being at least three times the size of the start-momentum. A smaller k-space leads to reflections of the wave packets and wrong results. To avoid that the wave packets leave the grid, we use a self-adapting co-moving grid, which allows us to reach propagation times beyond 100\,fs.
The electron emission from tips is not spin selective, therefore, we end up with three antisymmetric and one symmetric spatial wave function \cite{Lougovski2011}. The shown energy gap in Fig. \ref{fig:Figure4} is then the incoherent sum of all parts. For quantum statistics effects an overlap both in real and k-space is necessary. Coulomb repulsion, however, separates the wave function especially in k-space, which is why we do not observe any difference between antisymmetric and symmetric parts after 100\,fs propagation.

\end{appendices}

\bibliography{literature}


\begin{thebibliography}{49}
\ifx \bisbn   \undefined \def \bisbn  #1{ISBN #1}\fi
\ifx \binits  \undefined \def \binits#1{#1}\fi
\ifx \bauthor  \undefined \def \bauthor#1{#1}\fi
\ifx \batitle  \undefined \def \batitle#1{#1}\fi
\ifx \bjtitle  \undefined \def \bjtitle#1{#1}\fi
\ifx \bvolume  \undefined \def \bvolume#1{\textbf{#1}}\fi
\ifx \byear  \undefined \def \byear#1{#1}\fi
\ifx \bissue  \undefined \def \bissue#1{#1}\fi
\ifx \bfpage  \undefined \def \bfpage#1{#1}\fi
\ifx \blpage  \undefined \def \blpage #1{#1}\fi
\ifx \burl  \undefined \def \burl#1{\textsf{#1}}\fi
\ifx \doiurl  \undefined \def \doiurl#1{\url{https://doi.org/#1}}\fi
\ifx \betal  \undefined \def \betal{\textit{et al.}}\fi
\ifx \binstitute  \undefined \def \binstitute#1{#1}\fi
\ifx \binstitutionaled  \undefined \def \binstitutionaled#1{#1}\fi
\ifx \bctitle  \undefined \def \bctitle#1{#1}\fi
\ifx \beditor  \undefined \def \beditor#1{#1}\fi
\ifx \bpublisher  \undefined \def \bpublisher#1{#1}\fi
\ifx \bbtitle  \undefined \def \bbtitle#1{#1}\fi
\ifx \bedition  \undefined \def \bedition#1{#1}\fi
\ifx \bseriesno  \undefined \def \bseriesno#1{#1}\fi
\ifx \blocation  \undefined \def \blocation#1{#1}\fi
\ifx \bsertitle  \undefined \def \bsertitle#1{#1}\fi
\ifx \bsnm \undefined \def \bsnm#1{#1}\fi
\ifx \bsuffix \undefined \def \bsuffix#1{#1}\fi
\ifx \bparticle \undefined \def \bparticle#1{#1}\fi
\ifx \barticle \undefined \def \barticle#1{#1}\fi
\bibcommenthead
\ifx \bconfdate \undefined \def \bconfdate #1{#1}\fi
\ifx \botherref \undefined \def \botherref #1{#1}\fi
\ifx \url \undefined \def \url#1{\textsf{#1}}\fi
\ifx \bchapter \undefined \def \bchapter#1{#1}\fi
\ifx \bbook \undefined \def \bbook#1{#1}\fi
\ifx \bcomment \undefined \def \bcomment#1{#1}\fi
\ifx \oauthor \undefined \def \oauthor#1{#1}\fi
\ifx \citeauthoryear \undefined \def \citeauthoryear#1{#1}\fi
\ifx \endbibitem  \undefined \def \endbibitem {}\fi
\ifx \bconflocation  \undefined \def \bconflocation#1{#1}\fi
\ifx \arxivurl  \undefined \def \arxivurl#1{\textsf{#1}}\fi
\csname PreBibitemsHook\endcsname

%%% 1
\bibitem{Bloch2022}
\begin{barticle}
\bauthor{\bsnm{Bloch}, \binits{J.}},
\bauthor{\bsnm{Cavalleri}, \binits{A.}},
\bauthor{\bsnm{Galitski}, \binits{V.}},
\bauthor{\bsnm{Hafezi}, \binits{M.}},
\bauthor{\bsnm{Rubio}, \binits{A.}}:
\batitle{Strongly correlated electron{\textendash}photon systems}.
\bjtitle{Nature}
\bvolume{606}(\bissue{7912}),
\bfpage{41}--\blpage{48}
(\byear{2022}).
\doiurl{10.1038/s41586-022-04726-w}
\end{barticle}
\endbibitem

%%% 2
\bibitem{Kouzakov2003}
\begin{botherref}
\oauthor{\bsnm{Kouzakov}, \binits{K.A.}},
\oauthor{\bsnm{Berakdar}, \binits{J.}}:
Photoinduced emission of cooper pairs from superconductors.
Physical Review Letters
\textbf{91}(25)
(2003).
\doiurl{10.1103/physrevlett.91.257007}
\end{botherref}
\endbibitem

%%% 3
\bibitem{Sobota2021}
\begin{botherref}
\oauthor{\bsnm{Sobota}, \binits{J.A.}},
\oauthor{\bsnm{He}, \binits{Y.}},
\oauthor{\bsnm{Shen}, \binits{Z.-X.}}:
Angle-resolved photoemission studies of quantum materials.
Reviews of Modern Physics
\textbf{93}(2)
(2021).
\doiurl{10.1103/revmodphys.93.025006}
\end{botherref}
\endbibitem

%%% 4
\bibitem{Wang2013}
\begin{barticle}
\bauthor{\bsnm{Wang}, \binits{Y.H.}},
\bauthor{\bsnm{Steinberg}, \binits{H.}},
\bauthor{\bsnm{Jarillo-Herrero}, \binits{P.}},
\bauthor{\bsnm{Gedik}, \binits{N.}}:
\batitle{Observation of floquet-bloch states on the surface of a topological
  insulator}.
\bjtitle{Science}
\bvolume{342}(\bissue{6157}),
\bfpage{453}--\blpage{457}
(\byear{2013}).
\doiurl{10.1126/science.1239834}
\end{barticle}
\endbibitem

%%% 5
\bibitem{Wallauer2021}
\begin{barticle}
\bauthor{\bsnm{Wallauer}, \binits{R.}},
\bauthor{\bsnm{Raths}, \binits{M.}},
\bauthor{\bsnm{Stallberg}, \binits{K.}},
\bauthor{\bsnm{M\"{u}nster}, \binits{L.}},
\bauthor{\bsnm{Brandstetter}, \binits{D.}},
\bauthor{\bsnm{Yang}, \binits{X.}},
\bauthor{\bsnm{G\"{u}dde}, \binits{J.}},
\bauthor{\bsnm{Puschnig}, \binits{P.}},
\bauthor{\bsnm{Soubatch}, \binits{S.}},
\bauthor{\bsnm{Kumpf}, \binits{C.}},
\bauthor{\bsnm{Bocquet}, \binits{F.C.}},
\bauthor{\bsnm{Tautz}, \binits{F.S.}},
\bauthor{\bsnm{H\"{o}fer}, \binits{U.}}:
\batitle{Tracing orbital images on ultrafast time scales}.
\bjtitle{Science}
\bvolume{371}(\bissue{6533}),
\bfpage{1056}--\blpage{1059}
(\byear{2021}).
\doiurl{10.1126/science.abf3286}
\end{barticle}
\endbibitem

%%% 6
\bibitem{Johnson2022}
\begin{botherref}
\oauthor{\bsnm{Johnson}, \binits{C.W.}},
\oauthor{\bsnm{Schmid}, \binits{A.K.}},
\oauthor{\bsnm{Mankos}, \binits{M.}},
\oauthor{\bsnm{Röpke}, \binits{R.}},
\oauthor{\bsnm{Kerker}, \binits{N.}},
\oauthor{\bsnm{Wong}, \binits{E.K.}},
\oauthor{\bsnm{Ogletree}, \binits{D.F.}},
\oauthor{\bsnm{Minor}, \binits{A.M.}},
\oauthor{\bsnm{Stibor}, \binits{A.}}:
Near-monochromatic tuneable cryogenic niobium electron field emitter.
arXiv
(2022).
\doiurl{10.48550/ARXIV.2205.05767}.
\url{https://arxiv.org/abs/2205.05767}
\end{botherref}
\endbibitem

%%% 7
\bibitem{Glaeser2015}
\begin{barticle}
\bauthor{\bsnm{Glaeser}, \binits{R.M.}}:
\batitle{How good can cryo-{EM} become?}
\bjtitle{Nature Methods}
\bvolume{13}(\bissue{1}),
\bfpage{28}--\blpage{32}
(\byear{2015}).
\doiurl{10.1038/nmeth.3695}
\end{barticle}
\endbibitem

%%% 8
\bibitem{MagaaLoaiza2019}
\begin{barticle}
\bauthor{\bsnm{Maga{\~{n}}a-Loaiza}, \binits{O.S.}},
\bauthor{\bsnm{Boyd}, \binits{R.W.}}:
\batitle{Quantum imaging and information}.
\bjtitle{Reports on Progress in Physics}
\bvolume{82}(\bissue{12}),
\bfpage{124401}
(\byear{2019}).
\doiurl{10.1088/1361-6633/ab5005}
\end{barticle}
\endbibitem

%%% 9
\bibitem{Feist2017}
\begin{barticle}
\bauthor{\bsnm{Feist}, \binits{A.}},
\bauthor{\bsnm{Bach}, \binits{N.}},
\bauthor{\bparticle{da} \bsnm{Silva}, \binits{N.R.}},
\bauthor{\bsnm{Danz}, \binits{T.}},
\bauthor{\bsnm{M\"{o}ller}, \binits{M.}},
\bauthor{\bsnm{Priebe}, \binits{K.E.}},
\bauthor{\bsnm{Domr\"{o}se}, \binits{T.}},
\bauthor{\bsnm{Gatzmann}, \binits{J.G.}},
\bauthor{\bsnm{Rost}, \binits{S.}},
\bauthor{\bsnm{Schauss}, \binits{J.}},
\bauthor{\bsnm{Strauch}, \binits{S.}},
\bauthor{\bsnm{Bormann}, \binits{R.}},
\bauthor{\bsnm{Sivis}, \binits{M.}},
\bauthor{\bsnm{Sch\"{a}fer}, \binits{S.}},
\bauthor{\bsnm{Ropers}, \binits{C.}}:
\batitle{Ultrafast transmission electron microscopy using a laser-driven field
  emitter: Femtosecond resolution with a high coherence electron beam}.
\bjtitle{Ultramicroscopy}
\bvolume{176},
\bfpage{63}--\blpage{73}
(\byear{2017}).
\doiurl{10.1016/j.ultramic.2016.12.005}
\end{barticle}
\endbibitem

%%% 10
\bibitem{Arbouet2018}
\begin{bchapter}
\bauthor{\bsnm{Arbouet}, \binits{A.}},
\bauthor{\bsnm{Caruso}, \binits{G.M.}},
\bauthor{\bsnm{Houdellier}, \binits{F.}}:
\bctitle{Chapter one - ultrafast transmission electron microscopy: Historical
  development, instrumentation, and applications}.
In: \beditor{\bsnm{Hawkes}, \binits{P.W.}} (ed.)
\bbtitle{Advances in Imaging and Electron Physics}.
\bsertitle{Advances in Imaging and Electron Physics},
vol. \bseriesno{207},
pp. \bfpage{1}--\blpage{72}.
\bpublisher{Elsevier}, \blocation{???}
(\byear{2018}).
\doiurl{10.1016/bs.aiep.2018.06.001}.
\burl{https://www.sciencedirect.com/science/article/pii/S1076567018300223}
\end{bchapter}
\endbibitem

%%% 11
\bibitem{Zong2018}
\begin{barticle}
\bauthor{\bsnm{Zong}, \binits{A.}},
\bauthor{\bsnm{Kogar}, \binits{A.}},
\bauthor{\bsnm{Bie}, \binits{Y.-Q.}},
\bauthor{\bsnm{Rohwer}, \binits{T.}},
\bauthor{\bsnm{Lee}, \binits{C.}},
\bauthor{\bsnm{Baldini}, \binits{E.}},
\bauthor{\bsnm{Erge{\c{c}}en}, \binits{E.}},
\bauthor{\bsnm{Yilmaz}, \binits{M.B.}},
\bauthor{\bsnm{Freelon}, \binits{B.}},
\bauthor{\bsnm{Sie}, \binits{E.J.}},
\bauthor{\bsnm{Zhou}, \binits{H.}},
\bauthor{\bsnm{Straquadine}, \binits{J.}},
\bauthor{\bsnm{Walmsley}, \binits{P.}},
\bauthor{\bsnm{Dolgirev}, \binits{P.E.}},
\bauthor{\bsnm{Rozhkov}, \binits{A.V.}},
\bauthor{\bsnm{Fisher}, \binits{I.R.}},
\bauthor{\bsnm{Jarillo-Herrero}, \binits{P.}},
\bauthor{\bsnm{Fine}, \binits{B.V.}},
\bauthor{\bsnm{Gedik}, \binits{N.}}:
\batitle{Evidence for topological defects in a photoinduced phase transition}.
\bjtitle{Nature Physics}
\bvolume{15}(\bissue{1}),
\bfpage{27}--\blpage{31}
(\byear{2018}).
\doiurl{10.1038/s41567-018-0311-9}
\end{barticle}
\endbibitem

%%% 12
\bibitem{England2014}
\begin{barticle}
\bauthor{\bsnm{England}, \binits{R.J.}},
\bauthor{\bsnm{Noble}, \binits{R.J.}},
\bauthor{\bsnm{Bane}, \binits{K.}},
\bauthor{\bsnm{Dowell}, \binits{D.H.}},
\bauthor{\bsnm{Ng}, \binits{C.-K.}},
\bauthor{\bsnm{Spencer}, \binits{J.E.}},
\bauthor{\bsnm{Tantawi}, \binits{S.}},
\bauthor{\bsnm{Wu}, \binits{Z.}},
\bauthor{\bsnm{Byer}, \binits{R.L.}},
\bauthor{\bsnm{Peralta}, \binits{E.}},
\bauthor{\bsnm{Soong}, \binits{K.}},
\bauthor{\bsnm{Chang}, \binits{C.-M.}},
\bauthor{\bsnm{Montazeri}, \binits{B.}},
\bauthor{\bsnm{Wolf}, \binits{S.J.}},
\bauthor{\bsnm{Cowan}, \binits{B.}},
\bauthor{\bsnm{Dawson}, \binits{J.}},
\bauthor{\bsnm{Gai}, \binits{W.}},
\bauthor{\bsnm{Hommelhoff}, \binits{P.}},
\bauthor{\bsnm{Huang}, \binits{Y.-C.}},
\bauthor{\bsnm{Jing}, \binits{C.}},
\bauthor{\bsnm{McGuinness}, \binits{C.}},
\bauthor{\bsnm{Palmer}, \binits{R.B.}},
\bauthor{\bsnm{Naranjo}, \binits{B.}},
\bauthor{\bsnm{Rosenzweig}, \binits{J.}},
\bauthor{\bsnm{Travish}, \binits{G.}},
\bauthor{\bsnm{Mizrahi}, \binits{A.}},
\bauthor{\bsnm{Schachter}, \binits{L.}},
\bauthor{\bsnm{Sears}, \binits{C.}},
\bauthor{\bsnm{Werner}, \binits{G.R.}},
\bauthor{\bsnm{Yoder}, \binits{R.B.}}:
\batitle{Dielectric laser accelerators}.
\bjtitle{Rev. Mod. Phys.}
\bvolume{86},
\bfpage{1337}--\blpage{1389}
(\byear{2014}).
\doiurl{10.1103/RevModPhys.86.1337}
\end{barticle}
\endbibitem

%%% 13
\bibitem{Ludwig2019}
\begin{barticle}
\bauthor{\bsnm{Ludwig}, \binits{M.}},
\bauthor{\bsnm{Aguirregabiria}, \binits{G.}},
\bauthor{\bsnm{Ritzkowsky}, \binits{F.}},
\bauthor{\bsnm{Rybka}, \binits{T.}},
\bauthor{\bsnm{Marinica}, \binits{D.C.}},
\bauthor{\bsnm{Aizpurua}, \binits{J.}},
\bauthor{\bsnm{Borisov}, \binits{A.G.}},
\bauthor{\bsnm{Leitenstorfer}, \binits{A.}},
\bauthor{\bsnm{Brida}, \binits{D.}}:
\batitle{Sub-femtosecond electron transport in a nanoscale gap}.
\bjtitle{Nature Physics}
(\byear{2019}).
\doiurl{10.1038/s41567-019-0745-8}
\end{barticle}
\endbibitem

%%% 14
\bibitem{Hergert2021}
\begin{barticle}
\bauthor{\bsnm{Hergert}, \binits{G.}},
\bauthor{\bsnm{W\"{o}ste}, \binits{A.}},
\bauthor{\bsnm{Vogelsang}, \binits{J.}},
\bauthor{\bsnm{Quenzel}, \binits{T.}},
\bauthor{\bsnm{Wang}, \binits{D.}},
\bauthor{\bsnm{Gross}, \binits{P.}},
\bauthor{\bsnm{Lienau}, \binits{C.}}:
\batitle{Probing transient localized electromagnetic fields using low-energy
  point-projection electron microscopy}.
\bjtitle{{ACS} Photonics}
\bvolume{8}(\bissue{9}),
\bfpage{2573}--\blpage{2580}
(\byear{2021}).
\doiurl{10.1021/acsphotonics.1c00775}
\end{barticle}
\endbibitem

%%% 15
\bibitem{lHuillier1983}
\begin{barticle}
\bauthor{\bsnm{l{\textquotesingle}Huillier}, \binits{A.}},
\bauthor{\bsnm{Lompre}, \binits{L.A.}},
\bauthor{\bsnm{Mainfray}, \binits{G.}},
\bauthor{\bsnm{Manus}, \binits{C.}}:
\batitle{Multiply charged ions induced by multiphoton absorption in rare gases
  at 0.53 $\mu$m}.
\bjtitle{Physical Review A}
\bvolume{27}(\bissue{5}),
\bfpage{2503}--\blpage{2512}
(\byear{1983}).
\doiurl{10.1103/physreva.27.2503}
\end{barticle}
\endbibitem

%%% 16
\bibitem{Weber2000}
\begin{barticle}
\bauthor{\bsnm{Weber}, \binits{T.}},
\bauthor{\bsnm{Giessen}, \binits{H.}},
\bauthor{\bsnm{Weckenbrock}, \binits{M.}},
\bauthor{\bsnm{Urbasch}, \binits{G.}},
\bauthor{\bsnm{Staudte}, \binits{A.}},
\bauthor{\bsnm{Spielberger}, \binits{L.}},
\bauthor{\bsnm{Jagutzki}, \binits{O.}},
\bauthor{\bsnm{Mergel}, \binits{V.}},
\bauthor{\bsnm{Vollmer}, \binits{M.}},
\bauthor{\bsnm{D\"{o}rner}, \binits{R.}}:
\batitle{Correlated electron emission in multiphoton double ionization}.
\bjtitle{Nature}
\bvolume{405}(\bissue{6787}),
\bfpage{658}--\blpage{661}
(\byear{2000}).
\doiurl{10.1038/35015033}
\end{barticle}
\endbibitem

%%% 17
\bibitem{Becker2012}
\begin{barticle}
\bauthor{\bsnm{Becker}, \binits{W.}},
\bauthor{\bsnm{Liu}, \binits{X.}},
\bauthor{\bsnm{Ho}, \binits{P.J.}},
\bauthor{\bsnm{Eberly}, \binits{J.H.}}:
\batitle{Theories of photoelectron correlation in laser-driven multiple atomic
  ionization}.
\bjtitle{Rev. Mod. Phys.}
\bvolume{84},
\bfpage{1011}--\blpage{1043}
(\byear{2012}).
\doiurl{10.1103/RevModPhys.84.1011}
\end{barticle}
\endbibitem

%%% 18
\bibitem{Hommelhoff2006_1}
\begin{botherref}
\oauthor{\bsnm{Hommelhoff}, \binits{P.}},
\oauthor{\bsnm{Sortais}, \binits{Y.}},
\oauthor{\bsnm{Aghajani-Talesh}, \binits{A.}},
\oauthor{\bsnm{Kasevich}, \binits{M.A.}}:
Field emission tip as a nanometer source of free electron femtosecond pulses.
Physical Review Letters
\textbf{96}(7)
(2006).
\doiurl{10.1103/physrevlett.96.077401}
\end{botherref}
\endbibitem

%%% 19
\bibitem{Ropers2007}
\begin{barticle}
\bauthor{\bsnm{Ropers}, \binits{C.}},
\bauthor{\bsnm{Solli}, \binits{D.R.}},
\bauthor{\bsnm{Schulz}, \binits{C.P.}},
\bauthor{\bsnm{Lienau}, \binits{C.}},
\bauthor{\bsnm{Elsaesser}, \binits{T.}}:
\batitle{Localized multiphoton emission of femtosecond electron pulses from
  metal nanotips}.
\bjtitle{Phys. Rev. Lett.}
\bvolume{98}(\bissue{4}),
\bfpage{043907}
(\byear{2007}).
\doiurl{10.1103/physrevlett.98.043907}
\end{barticle}
\endbibitem

%%% 20
\bibitem{Bormann2010}
\begin{barticle}
\bauthor{\bsnm{Bormann}, \binits{R.}},
\bauthor{\bsnm{Gulde}, \binits{M.}},
\bauthor{\bsnm{Weismann}, \binits{A.}},
\bauthor{\bsnm{Yalunin}, \binits{S.V.}},
\bauthor{\bsnm{Ropers}, \binits{C.}}:
\batitle{Tip-enhanced strong-field photoemission}.
\bjtitle{Phys. Rev. Lett.}
\bvolume{105}(\bissue{14}),
\bfpage{147601}
(\byear{2010}).
\doiurl{10.1103/physrevlett.105.147601}
\end{barticle}
\endbibitem

%%% 21
\bibitem{Krger2018}
\begin{barticle}
\bauthor{\bsnm{Kr\"{u}ger}, \binits{M.}},
\bauthor{\bsnm{Lemell}, \binits{C.}},
\bauthor{\bsnm{Wachter}, \binits{G.}},
\bauthor{\bsnm{Burgdoerfer}, \binits{J.}},
\bauthor{\bsnm{Hommelhoff}, \binits{P.}}:
\batitle{Attosecond physics phenomena at nanometric tips}.
\bjtitle{Journal of Physics B: Atomic, Molecular and Optical Physics}
\bvolume{51}(\bissue{17}),
\bfpage{172001}
(\byear{2018}).
\doiurl{10.1088/1361-6455/aac6ac}
\end{barticle}
\endbibitem

%%% 22
\bibitem{Thomas2013}
\begin{barticle}
\bauthor{\bsnm{Thomas}, \binits{S.}},
\bauthor{\bsnm{Kr\"{u}ger}, \binits{M.}},
\bauthor{\bsnm{F\"{o}rster}, \binits{M.}},
\bauthor{\bsnm{Schenk}, \binits{M.}},
\bauthor{\bsnm{Hommelhoff}, \binits{P.}}:
\batitle{Probing of optical near-fields by electron rescattering on the 1 nm
  scale}.
\bjtitle{Nano Letters}
\bvolume{13}(\bissue{10}),
\bfpage{4790}--\blpage{4794}
(\byear{2013}).
\doiurl{10.1021/nl402407r}
\end{barticle}
\endbibitem

%%% 23
\bibitem{Jiang2018_2}
\begin{barticle}
\bauthor{\bsnm{Jiang}, \binits{Y.}},
\bauthor{\bsnm{Chen}, \binits{Z.}},
\bauthor{\bsnm{Han}, \binits{Y.}},
\bauthor{\bsnm{Deb}, \binits{P.}},
\bauthor{\bsnm{Gao}, \binits{H.}},
\bauthor{\bsnm{Xie}, \binits{S.}},
\bauthor{\bsnm{Purohit}, \binits{P.}},
\bauthor{\bsnm{Tate}, \binits{M.W.}},
\bauthor{\bsnm{Park}, \binits{J.}},
\bauthor{\bsnm{Gruner}, \binits{S.M.}},
\bauthor{\bsnm{Elser}, \binits{V.}},
\bauthor{\bsnm{Muller}, \binits{D.A.}}:
\batitle{Electron ptychography of 2d materials to deep sub-{\aa}ngstr\"{o}m
  resolution}.
\bjtitle{Nature}
\bvolume{559}(\bissue{7714}),
\bfpage{343}--\blpage{349}
(\byear{2018}).
\doiurl{10.1038/s41586-018-0298-5}
\end{barticle}
\endbibitem

%%% 24
\bibitem{Bach2019}
\begin{barticle}
\bauthor{\bsnm{Bach}, \binits{N.}},
\bauthor{\bsnm{Domr\"{o}se}, \binits{T.}},
\bauthor{\bsnm{Feist}, \binits{A.}},
\bauthor{\bsnm{Rittmann}, \binits{T.}},
\bauthor{\bsnm{Strauch}, \binits{S.}},
\bauthor{\bsnm{Ropers}, \binits{C.}},
\bauthor{\bsnm{Sch\"{a}fer}, \binits{S.}}:
\batitle{Coulomb interactions in high-coherence femtosecond electron pulses
  from tip emitters}.
\bjtitle{Structural Dynamics}
\bvolume{6}(\bissue{1}),
\bfpage{014301}
(\byear{2019}).
\doiurl{10.1063/1.5066093}
\end{barticle}
\endbibitem

%%% 25
\bibitem{Tsarev2021}
\begin{barticle}
\bauthor{\bsnm{Tsarev}, \binits{M.}},
\bauthor{\bsnm{Ryabov}, \binits{A.}},
\bauthor{\bsnm{Baum}, \binits{P.}}:
\batitle{Measurement of temporal coherence of free electrons by time-domain
  electron interferometry}.
\bjtitle{Phys. Rev. Lett.}
\bvolume{127},
\bfpage{165501}
(\byear{2021}).
\doiurl{10.1103/PhysRevLett.127.165501}
\end{barticle}
\endbibitem

%%% 26
\bibitem{Kuwahara2021}
\begin{barticle}
\bauthor{\bsnm{Kuwahara}, \binits{M.}},
\bauthor{\bsnm{Yoshida}, \binits{Y.}},
\bauthor{\bsnm{Nagata}, \binits{W.}},
\bauthor{\bsnm{Nakakura}, \binits{K.}},
\bauthor{\bsnm{Furui}, \binits{M.}},
\bauthor{\bsnm{Ishida}, \binits{T.}},
\bauthor{\bsnm{Saitoh}, \binits{K.}},
\bauthor{\bsnm{Ujihara}, \binits{T.}},
\bauthor{\bsnm{Tanaka}, \binits{N.}}:
\batitle{Intensity interference in a coherent spin-polarized electron beam}.
\bjtitle{Physical Review Letters}
\bvolume{126}(\bissue{12}),
\bfpage{125501}
(\byear{2021}).
\doiurl{10.1103/physrevlett.126.125501}
\end{barticle}
\endbibitem

%%% 27
\bibitem{Yudin2001}
\begin{botherref}
\oauthor{\bsnm{Yudin}, \binits{G.L.}},
\oauthor{\bsnm{Ivanov}, \binits{M.Y.}}:
Nonadiabatic tunnel ionization: Looking inside a laser cycle.
Physical Review A
\textbf{64}(1)
(2001).
\doiurl{10.1103/physreva.64.013409}
\end{botherref}
\endbibitem

%%% 28
\bibitem{Keramati2021}
\begin{barticle}
\bauthor{\bsnm{Keramati}, \binits{S.}},
\bauthor{\bsnm{Brunner}, \binits{W.}},
\bauthor{\bsnm{Gay}, \binits{T.J.}},
\bauthor{\bsnm{Batelaan}, \binits{H.}}:
\batitle{Non-poissonian ultrashort nanoscale electron pulses}.
\bjtitle{Phys. Rev. Lett.}
\bvolume{127},
\bfpage{180602}
(\byear{2021}).
\doiurl{10.1103/PhysRevLett.127.180602}
\end{barticle}
\endbibitem

%%% 29
\bibitem{berchera2019quantum}
\begin{barticle}
\bauthor{\bsnm{Berchera}, \binits{I.R.}},
\bauthor{\bsnm{Degiovanni}, \binits{I.P.}}:
\batitle{Quantum imaging with sub-poissonian light: challenges and perspectives
  in optical metrology}.
\bjtitle{Metrologia}
\bvolume{56}(\bissue{2}),
\bfpage{024001}
(\byear{2019})
\end{barticle}
\endbibitem

%%% 30
\bibitem{Zrenner2002}
\begin{barticle}
\bauthor{\bsnm{Zrenner}, \binits{A.}},
\bauthor{\bsnm{Beham}, \binits{E.}},
\bauthor{\bsnm{Stufler}, \binits{S.}},
\bauthor{\bsnm{Findeis}, \binits{F.}},
\bauthor{\bsnm{Bichler}, \binits{M.}},
\bauthor{\bsnm{Abstreiter}, \binits{G.}}:
\batitle{Coherent properties of a two-level system based on a quantum-dot
  photodiode}.
\bjtitle{Nature}
\bvolume{418}(\bissue{6898}),
\bfpage{612}--\blpage{614}
(\byear{2002}).
\doiurl{10.1038/nature00912}
\end{barticle}
\endbibitem

%%% 31
\bibitem{Hommelhoff2006_2}
\begin{barticle}
\bauthor{\bsnm{Hommelhoff}, \binits{P.}},
\bauthor{\bsnm{Sortais}, \binits{Y.}},
\bauthor{\bsnm{Aghajani-Talesh}, \binits{A.}},
\bauthor{\bsnm{Kasevich}, \binits{M.A.}}:
\batitle{Field emission tip as a nanometer source of free electron femtosecond
  pulses}.
\bjtitle{Physical Review Letters}
\bvolume{96}(\bissue{7}),
\bfpage{077401}
(\byear{2006}).
\doiurl{10.1103/physrevlett.96.077401}
\end{barticle}
\endbibitem

%%% 32
\bibitem{Duchet2021}
\begin{barticle}
\bauthor{\bsnm{Duchet}, \binits{M.}},
\bauthor{\bsnm{Perisanu}, \binits{S.}},
\bauthor{\bsnm{Purcell}, \binits{S.T.}},
\bauthor{\bsnm{Constant}, \binits{E.}},
\bauthor{\bsnm{Loriot}, \binits{V.}},
\bauthor{\bsnm{Yanagisawa}, \binits{H.}},
\bauthor{\bsnm{Kling}, \binits{M.F.}},
\bauthor{\bsnm{Lepine}, \binits{F.}},
\bauthor{\bsnm{Ayari}, \binits{A.}}:
\batitle{Femtosecond laser induced resonant tunneling in an individual quantum
  dot attached to a nanotip}.
\bjtitle{ACS Photonics}
\bvolume{8}(\bissue{2}),
\bfpage{505}--\blpage{511}
(\byear{2021})
{\href{https://arxiv.org/abs/https://doi.org/10.1021/acsphotonics.0c01490}{{https://doi.org/10.1021/acsphotonics.0c01490}}}.
\doiurl{10.1021/acsphotonics.0c01490}
\end{barticle}
\endbibitem

%%% 33
\bibitem{Krger2011}
\begin{barticle}
\bauthor{\bsnm{Kr\"{u}ger}, \binits{M.}},
\bauthor{\bsnm{Schenk}, \binits{M.}},
\bauthor{\bsnm{Hommelhoff}, \binits{P.}}:
\batitle{Attosecond control of electrons emitted from a nanoscale metal tip}.
\bjtitle{Nature}
\bvolume{475}(\bissue{7354}),
\bfpage{78}--\blpage{81}
(\byear{2011}).
\doiurl{10.1038/nature10196}
\end{barticle}
\endbibitem

%%% 34
\bibitem{Herink2012}
\begin{barticle}
\bauthor{\bsnm{Herink}, \binits{G.}},
\bauthor{\bsnm{Solli}, \binits{D.R.}},
\bauthor{\bsnm{Gulde}, \binits{M.}},
\bauthor{\bsnm{Ropers}, \binits{C.}}:
\batitle{Field-driven photoemission from nanostructures quenches the quiver
  motion}.
\bjtitle{Nature}
\bvolume{483}(\bissue{7388}),
\bfpage{190}--\blpage{193}
(\byear{2012}).
\doiurl{10.1038/nature10878}
\end{barticle}
\endbibitem

%%% 35
\bibitem{Paulus1994}
\begin{barticle}
\bauthor{\bsnm{Paulus}, \binits{G.G.}},
\bauthor{\bsnm{Becker}, \binits{W.}},
\bauthor{\bsnm{Nicklich}, \binits{W.}},
\bauthor{\bsnm{Walther}, \binits{H.}}:
\batitle{Rescattering effects in above-threshold ionization: a classical
  model}.
\bjtitle{Journal of Physics B: Atomic, Molecular and Optical Physics}
\bvolume{27}(\bissue{21}),
\bfpage{703}--\blpage{708}
(\byear{1994}).
\doiurl{10.1088/0953-4075/27/21/003}
\end{barticle}
\endbibitem

%%% 36
\bibitem{Corkum1993}
\begin{barticle}
\bauthor{\bsnm{Corkum}, \binits{P.B.}}:
\batitle{Plasma perspective on strong field multiphoton ionization}.
\bjtitle{Phys. Rev. Lett.}
\bvolume{71},
\bfpage{1994}--\blpage{1997}
(\byear{1993}).
\doiurl{10.1103/PhysRevLett.71.1994}
\end{barticle}
\endbibitem

%%% 37
\bibitem{Lewenstein1994}
\begin{barticle}
\bauthor{\bsnm{Lewenstein}, \binits{M.}},
\bauthor{\bsnm{Balcou}, \binits{P.}},
\bauthor{\bsnm{Ivanov}, \binits{M.Y.}},
\bauthor{\bsnm{L'Huillier}, \binits{A.}},
\bauthor{\bsnm{Corkum}, \binits{P.B.}}:
\batitle{Theory of high-harmonic generation by low-frequency laser fields}.
\bjtitle{Physical Review A}
\bvolume{49}(\bissue{3}),
\bfpage{2117}--\blpage{2132}
(\byear{1994}).
\doiurl{10.1103/physreva.49.2117}
\end{barticle}
\endbibitem

%%% 38
\bibitem{Camus2012}
\begin{barticle}
\bauthor{\bsnm{Camus}, \binits{N.}},
\bauthor{\bsnm{Fischer}, \binits{B.}},
\bauthor{\bsnm{Kremer}, \binits{M.}},
\bauthor{\bsnm{Sharma}, \binits{V.}},
\bauthor{\bsnm{Rudenko}, \binits{A.}},
\bauthor{\bsnm{Bergues}, \binits{B.}},
\bauthor{\bsnm{K\"ubel}, \binits{M.}},
\bauthor{\bsnm{Johnson}, \binits{N.G.}},
\bauthor{\bsnm{Kling}, \binits{M.F.}},
\bauthor{\bsnm{Pfeifer}, \binits{T.}},
\bauthor{\bsnm{Ullrich}, \binits{J.}},
\bauthor{\bsnm{Moshammer}, \binits{R.}}:
\batitle{Attosecond correlated dynamics of two electrons passing through a
  transition state}.
\bjtitle{Phys. Rev. Lett.}
\bvolume{108},
\bfpage{073003}
(\byear{2012}).
\doiurl{10.1103/PhysRevLett.108.073003}
\end{barticle}
\endbibitem

%%% 39
\bibitem{Rudenko2007}
\begin{botherref}
\oauthor{\bsnm{Rudenko}, \binits{A.}},
\oauthor{\bparticle{de} \bsnm{Jesus}, \binits{V.L.B.}},
\oauthor{\bsnm{Ergler}, \binits{T.}},
\oauthor{\bsnm{Zrost}, \binits{K.}},
\oauthor{\bsnm{Feuerstein}, \binits{B.}},
\oauthor{\bsnm{Schr\"{o}ter}, \binits{C.D.}},
\oauthor{\bsnm{Moshammer}, \binits{R.}},
\oauthor{\bsnm{Ullrich}, \binits{J.}}:
Correlated two-electron momentum spectra for strong-field nonsequential double
  ionization of he at 800 nm.
Physical Review Letters
\textbf{99}(26)
(2007).
\doiurl{10.1103/physrevlett.99.263003}
\end{botherref}
\endbibitem

%%% 40
\bibitem{Staudte2007}
\begin{barticle}
\bauthor{\bsnm{Staudte}, \binits{A.}},
\bauthor{\bsnm{Ruiz}, \binits{C.}},
\bauthor{\bsnm{Sch\"offler}, \binits{M.}},
\bauthor{\bsnm{Sch\"ossler}, \binits{S.}},
\bauthor{\bsnm{Zeidler}, \binits{D.}},
\bauthor{\bsnm{Weber}, \binits{T.}},
\bauthor{\bsnm{Meckel}, \binits{M.}},
\bauthor{\bsnm{Villeneuve}, \binits{D.M.}},
\bauthor{\bsnm{Corkum}, \binits{P.B.}},
\bauthor{\bsnm{Becker}, \binits{A.}},
\bauthor{\bsnm{D\"orner}, \binits{R.}}:
\batitle{Binary and recoil collisions in strong field double ionization of
  helium}.
\bjtitle{Phys. Rev. Lett.}
\bvolume{99},
\bfpage{263002}
(\byear{2007}).
\doiurl{10.1103/PhysRevLett.99.263002}
\end{barticle}
\endbibitem

%%% 41
\bibitem{Kiesel2002}
\begin{barticle}
\bauthor{\bsnm{Kiesel}, \binits{H.}},
\bauthor{\bsnm{Renz}, \binits{A.}},
\bauthor{\bsnm{Hasselbach}, \binits{F.}}:
\batitle{Observation of hanbury brown{\textendash}twiss anticorrelations for
  free electrons}.
\bjtitle{Nature}
\bvolume{418}(\bissue{6896}),
\bfpage{392}--\blpage{394}
(\byear{2002}).
\doiurl{10.1038/nature00911}
\end{barticle}
\endbibitem

%%% 42
\bibitem{Kodama2011}
\begin{barticle}
\bauthor{\bsnm{Kodama}, \binits{T.}},
\bauthor{\bsnm{Osakabe}, \binits{N.}},
\bauthor{\bsnm{Tonomura}, \binits{A.}}:
\batitle{Correlation in a coherent electron beam}.
\bjtitle{Physical Review A}
\bvolume{83}(\bissue{6}),
\bfpage{063616}
(\byear{2011}).
\doiurl{10.1103/physreva.83.063616}
\end{barticle}
\endbibitem

%%% 43
\bibitem{Damm2011_phd}
\begin{botherref}
\oauthor{\bsnm{Damm}, \binits{A.}}:
Untersuchung der elektronendynamik von si(111) 7×7 und entwicklung eines
  flugzeitspektrometers für die zeit- und winkelaufgelöste
  zweiphotonen-photoemission.
PhD thesis,
Universität Marburg
(2011)
\end{botherref}
\endbibitem

%%% 44
\bibitem{Jagutzki2002_2}
\begin{barticle}
\bauthor{\bsnm{Jagutzki}, \binits{O.}},
\bauthor{\bsnm{Cerezo}, \binits{A.}},
\bauthor{\bsnm{Czasch}, \binits{A.}},
\bauthor{\bsnm{Dorner}, \binits{R.}},
\bauthor{\bsnm{Hattas}, \binits{M.}},
\bauthor{\bsnm{Huang}, \binits{M.}},
\bauthor{\bsnm{Mergel}, \binits{V.}},
\bauthor{\bsnm{Spillmann}, \binits{U.}},
\bauthor{\bsnm{Ullmann-Pfleger}, \binits{K.}},
\bauthor{\bsnm{Weber}, \binits{T.}},
\bauthor{\bsnm{Schmidt-Bocking}, \binits{H.}},
\bauthor{\bsnm{Smith}, \binits{G.D.W.}}:
\batitle{Multiple hit readout of a microchannel plate detector with a
  three-layer delay-line anode}.
\bjtitle{IEEE Transactions on Nuclear Science}
\bvolume{49}(\bissue{5}),
\bfpage{2477}--\blpage{2483}
(\byear{2002}).
\doiurl{10.1109/TNS.2002.803889}
\end{barticle}
\endbibitem

%%% 45
\bibitem{Fehre2018}
\begin{barticle}
\bauthor{\bsnm{Fehre}, \binits{K.}},
\bauthor{\bsnm{Trojanowskaja}, \binits{D.}},
\bauthor{\bsnm{Gatzke}, \binits{J.}},
\bauthor{\bsnm{Kunitski}, \binits{M.}},
\bauthor{\bsnm{Trinter}, \binits{F.}},
\bauthor{\bsnm{Zeller}, \binits{S.}},
\bauthor{\bsnm{Schmidt}, \binits{L.P.H.}},
\bauthor{\bsnm{Stohner}, \binits{J.}},
\bauthor{\bsnm{Berger}, \binits{R.}},
\bauthor{\bsnm{Czasch}, \binits{A.}},
\bauthor{\bsnm{Jagutzki}, \binits{O.}},
\bauthor{\bsnm{Jahnke}, \binits{T.}},
\bauthor{\bsnm{D\"{o}rner}, \binits{R.}},
\bauthor{\bsnm{Sch\"{o}ffler}, \binits{M.S.}}:
\batitle{Absolute ion detection efficiencies of microchannel plates and funnel
  microchannel plates for multi-coincidence detection}.
\bjtitle{Review of Scientific Instruments}
\bvolume{89}(\bissue{4}),
\bfpage{045112}
(\byear{2018}).
\doiurl{10.1063/1.5022564}
\end{barticle}
\endbibitem

%%% 46
\bibitem{Seiffert2018}
\begin{barticle}
\bauthor{\bsnm{Seiffert}, \binits{L.}},
\bauthor{\bsnm{Paschen}, \binits{T.}},
\bauthor{\bsnm{Hommelhoff}, \binits{P.}},
\bauthor{\bsnm{Fennel}, \binits{T.}}:
\batitle{High-order above-threshold photoemission from nanotips controlled with
  two-color laser fields}.
\bjtitle{Journal of Physics B: Atomic, Molecular and Optical Physics}
\bvolume{51}(\bissue{13}),
\bfpage{134001}
(\byear{2018}).
\doiurl{10.1088/1361-6455/aac34f}
\end{barticle}
\endbibitem

%%% 47
\bibitem{Krger2012_2}
\begin{barticle}
\bauthor{\bsnm{Kr\"{u}ger}, \binits{M.}},
\bauthor{\bsnm{Schenk}, \binits{M.}},
\bauthor{\bsnm{Hommelhoff}, \binits{P.}},
\bauthor{\bsnm{Wachter}, \binits{G.}},
\bauthor{\bsnm{Lemell}, \binits{C.}},
\bauthor{\bsnm{Burgd\"{o}rfer}, \binits{J.}}:
\batitle{Interaction of ultrashort laser pulses with metal nanotips: a model
  system for strong-field phenomena}.
\bjtitle{New Journal of Physics}
\bvolume{14}(\bissue{8}),
\bfpage{085019}
(\byear{2012}).
\doiurl{10.1088/1367-2630/14/8/085019}
\end{barticle}
\endbibitem

%%% 48
\bibitem{Maestri2000}
\begin{barticle}
\bauthor{\bsnm{Maestri}, \binits{J.J.V.}},
\bauthor{\bsnm{Landau}, \binits{R.H.}},
\bauthor{\bsnm{P{\'{a}}ez}, \binits{M.J.}}:
\batitle{Two-particle schr\"{o}dinger equation animations of wave
  packet{\textendash}wave packet scattering}.
\bjtitle{American Journal of Physics}
\bvolume{68}(\bissue{12}),
\bfpage{1113}--\blpage{1119}
(\byear{2000}).
\doiurl{10.1119/1.1286310}
\end{barticle}
\endbibitem

%%% 49
\bibitem{Lougovski2011}
\begin{botherref}
\oauthor{\bsnm{Lougovski}, \binits{P.}},
\oauthor{\bsnm{Batelaan}, \binits{H.}}:
Quantum description and properties of electrons emitted from pulsed nanotip
  electron sources.
Physical Review A
\textbf{84}(2)
(2011).
\doiurl{10.1103/physreva.84.023417}
\end{botherref}
\endbibitem

\end{thebibliography}

\subsection*{Acknowledgements}
The authors would like to thank Dr. Achim Czasch for technical discussions on the delay-line detector and Philip Dienstbier for discussions on the semi-classical simulation. This research was supported by the European Research Council (Consolidator Grant NearFieldAtto and Advanced Grant AccelOnChip) and the Deutsche Forschungsgemeinschaft (DFG, German Research Foundation) – Project-ID 429529648 – TRR 306 QuCoLiMa ("Quantum Cooperativity of Light and Matter") and Sonderforschungsbereich 953 ("Synthetic Carbon Allotropes"), Project-ID 182849149. J.H acknowledges funding from the Max Planck School of Photonics.
\subsection*{Author contributions}
S.M. and J.H. performed the experiment, analysed the data and generated the plots. S.M. performed the semi-classical simulations, J.H. the quantum-mechanical simulations. All authors wrote the manuscript.
\subsection*{Additional information}

\end{document}